\documentclass[aps,reprint,twocolumn,showpacs,preprintnumbers,amsmath,amssymb,nofootinbib,superscriptaddress,showkeys]{revtex4-1}

\usepackage{epsfig}
\usepackage{slashed}
\usepackage{color}

\usepackage[utf8]{inputenc}

\begin{document}

\title{Reexamining the perturbative renormalizability of coupled triplets}
	\author{Manuel Pavon Valderrama}\email{mpavon@buaa.edu.cn}
\affiliation{School of Physics, \\
Beihang University, Beijing 100191, China} 
\date{\today}

\begin{abstract} 
  \rule{0ex}{3ex}
  I reexamine the perturbative renormalizability of chiral two-pion exchange
  in two-nucleon scattering for coupled triplets when one-pion exchange
  has been fully iterated at leading order.
  Improving over previous works, it is shown that only two counterterms
  are required to obtain cutoff independent results, which is one less
  than in naive dimensional analysis.
  The explanation for this reduction is the existence of an attractive and
  repulsive eigenchannel in the one-pion exchange potential
  for the coupled triplets: the attractive eigenchannel can be renormalized
  like a regular attractive uncoupled triplet, while the repulsive
  eigenchannel is always finite regardless of
  whether there are counterterms or not.
  I discuss the implications of this finding for the power counting of
  the $^3S_1$-$^3D_1$ and $^3P_2$-$^3F_2$ partial waves.
\end{abstract}

\maketitle

\section{Introduction}
\label{sec:intro}

Effective field theories (EFTs) are model-independent and systematic
descriptions of low energy phenomena~\cite{Georgi:1993mps}.
They are particularly useful in situations where the formulation of
a more fundamental, high energy description is impossible or
impractical.
One of such use cases is nuclear physics~\cite{Machleidt:2017vls},
in which quantum chromodynamics (QCD), the fundamental theory of strong
interactions, is not solvable (except by brute
force~\cite{Beane:2010em,Aoki:2012tk})
at the low energy scales that are typical of nuclei
or nuclear processes.

In nuclear EFT~\cite{Epelbaum:2008ga,Machleidt:2011zz,Hammer:2019poc,vanKolck:2020llt,Machleidt:2024bwl}
the potential (or any other physical quantity for that matter)
is expanded as a power series
\begin{eqnarray}
  V_{\rm EFT} = \sum_{\nu = \nu_{\rm min}} V^{(\nu)} =
  \sum_{\nu = \nu_{\rm min}} \hat{V}^{(\nu)}\,{\left( \frac{Q}{M} \right)}^{\nu} \, ,
\end{eqnarray}
where the expansion parameter is a ratio of scales, $Q/M$, with $Q$ a
characteristic low energy scale such as the pion mass $m_{\pi}$ or
the momenta of nucleons within nuclei, and $M$ a high energy scale
such as the rho meson or nucleon masses, $m_{\rho}$ or $M_N$.

Provided $Q < M$, the EFT series will converge, at least at low orders:
chances are that the EFT expansion is actually an asymptotic expansion
(instead of a power series), in which case its divergent nature will
become manifest at high enough orders.
Be it as it may, each new term is increasingly difficult to calculate and
the expansion must be eventually truncated at a given order
$\nu = \nu_{\rm max}$, which induces a relative truncation
error of size $(Q/M)^{\nu_{\rm max} + 1}$.
The rules by which one decides the order at which a particular
contribution to the potential enters in the previous EFT expansion
are called power counting.
This set of rules is not unique and there is the question of whether
power counting should be grounded in naive dimensional
analysis (NDA)~\cite{Weinberg:1990rz,Weinberg:1991um} or
renormalizability~\cite{Nogga:2005hy,Birse:2005um,Valderrama:2009ei,Valderrama:2011mv,Long:2011qx,Long:2011xw,Long:2012ve}, which will be briefly
explained in the following paragraphs (and which is by itself
also the subject of a heated debate in the
community~\cite{Epelbaum:2006pt,Epelbaum:2009sd,Epelbaum:2018zli,Valderrama:2019yiv,Epelbaum:2020maf,Griesshammer:2021zzz,Gasparyan:2023rtj}).

Calculations in nuclear EFT can be done by iterating the EFT potential (expanded
up to the desired order) in the Schr\"odinger or Lippmann-Schwinger equation.
It is advisable (though, {\it prima facie}, not mandatory) to reexpand
calculations as to ensure that observables can be expressed
as a power series in terms of the power counting.
That is, to only treat non-perturbatively the {\it leading order} (${\rm LO}$)
part of the EFT potential, while subleading order contributions
enter as perturbative corrections.
Thus, {\it ideally}, the tool of choice of nuclear EFT would be distorted
wave perturbation theory, where the stress is in ideally as this is
technically challenging (if not outright prohibitively difficult)
in most settings, which explains why the most phenomenologically
successful EFT potentials are meant to be used
non-perturbatively~\cite{Epelbaum:2014efa,Epelbaum:2014sza,Entem:2017gor,Reinert:2017usi,Saha:2022oep}.

There is a crucial complication though that is rooted in the fact that
the EFT expansion is a long range (i.e. low energy) expansion,
generating potentials which exhibit the following power-law
behavior in coordinate space
\begin{eqnarray}
  V^{(\nu)}(\vec{r}\,) \propto \frac{1}{r^{\nu + 3}} \, ,
\end{eqnarray}
where $\nu$ refers to its order in NDA.
At long distances this is a feature: it signals convergence.
But at short distances the contributions to the EFT potential are
increasingly singular, requiring regularization and
renormalization.
It turns out that the renormalization of purely non-perturbative calculations
seems to be incompatible with power counting~\cite{PavonValderrama:2005wv,PavonValderrama:2005uj,Entem:2007jg,PavonValderrama:2007nu}.
In contrast, perturbative calculations (or, more precisely, a non-perturbative
leading order with perturbative subleading corrections, which is what it will
be meant by ``perturbative calculation'' when used without qualifiers
from now on) are renormalizable in a way that is compatible
with a sensible power counting.

Thus, if one is interested in the renormalizability of nuclear EFT
(understood as being able to prove that the cutoff can be removed),
it is mandatory to treat the subleading corrections perturbatively.
This is what was done in Refs.~\cite{Valderrama:2009ei,Valderrama:2011mv},
where it was shown that chiral two-pion exchange in coupled triplets
was renormalizable with six counterterms (i.e. by including six independent
contact-range operators in the calculations, or more generally by setting
six independent renormalization conditions).
Later, Long and Yang~\cite{Long:2011qx,Long:2011xw} reduced the aforementioned
number of counterterms from six to three in the coupled triplets,
which incidentally coincides with what is expected in NDA.
However Gasparyan and Epelbaum~\cite{Gasparyan:2022isg} demonstrated
that a large family of regulators do not generate cutoff independent
amplitudes in perturbative calculations, including the momentum space
calculations with separable regulators originally used
in Refs.~\cite{Long:2011qx,Long:2011xw}.
The significance of this interesting result has been recently
discussed in~\cite{Peng:2024aiz,Peng:2025ykg,Yang:2024yqv,PavonValderrama:2025tmp},
where it is apparent on the one hand that there are valid EFT arguments
by which this problem can be
averted~\cite{Peng:2024aiz,Peng:2025ykg,Yang:2024yqv} and
on the other that there exists regulators for which the amplitudes
are cutoff independent~\cite{Yang:2024yqv,PavonValderrama:2025tmp}.
But despite the previous issues,
and owing to the fact that coupled triplets can actually be renormalized
with two counterterms (which is the subject of the present manuscript),
Refs.~\cite{Long:2011qx,Long:2011xw} were actually correct about
how to renormalize the triplets (regardless of whether
their execution contained the subtle flaw discovered
in~\cite{Gasparyan:2022isg}).
In contrast,
coordinate space calculations with a sharp cutoff radius $R_c$ can be
easily proved to have a well-defined $R_c \to 0$ limit from a series of
semi-analytic arguments involving the detailed behavior of
the wave function at short distances~\cite{Valderrama:2009ei,Valderrama:2011mv},
though no explicit calculation of the cutoff dependence of the phase
shifts was originally provided there.
Recently, in~\cite{PavonValderrama:2025tmp} I have explicitly calculated
the cutoff dependence of the $^3P_0$ phase shifts, where it is shown
the absence of exceptional cutoffs for a series of coordinate
space regulators (including the original methods and
codes of Refs.~\cite{Valderrama:2009ei,Valderrama:2011mv}).

In the present manuscript, I further reduce the number of required counterterms
to renormalized the attractive triplets from six (or three) to two.
Though the explicit proof is rather technical, the basic idea of
why this is the case is easy to understand:
the perturbative renormalizability of two-pion exchange (TPE)
in the uncoupled triplets requires two or zero counterterms depending
on whether one is dealing with a triplet where the one-pion exchange (OPE)
tensor force is attractive or repulsive.
In coupled triplets the OPE tensor force is a matrix containing
an attractive and repulsive eigenvalue.
By renormalizing the coupled triplets in the eigenchannels where the sign of
the tensor force is well-defined (instead of the spectroscopic channels
corresponding to well-defined angular momenta),
it becomes apparent that two counterterm are enough to renormalize TPE
in said eigenchannels (provided the non-diagonal, interference term
does not require additional counterterm,
which happens to be the case).
This will be explained in detail and shown explicitly with concrete
calculations of the cutoff dependence of the phase shifts of
the $^3S_1$-$^3D_1$ and $^3P_2$-$^3F_2$ channels.

The manuscript is organized as follows: in Section \ref{sec:ren}
the perturbative renormalizability of the uncoupled and coupled
triplets is first reviewed (as it was originally used
in Refs.~\cite{Valderrama:2009ei,Valderrama:2011mv}), and then further
particularized to the attractive and repulsive
eigenchannels of the coupled triplets.
In Section \ref{sec:reg-and-ren} the $^3S_1$-$^3D_1$ and $^3P_2$-$^3F_2$
phase shifts and their cutoff dependence are calculated.
Finally, I summarize and discuss the previous findings
in Section \ref{sec:conclusions}, where the stress will be on what are the
power counting implications of the present results.

\section{Perturbative renormalizability}
\label{sec:ren}

In this Section, I will explain the distorted wave perturbation theory
formalism that is necessary for the calculation of
subleading order corrections.
The divergences that might appear during this process are carefully
analyzed and then it is explained how to renormalize them.
First I present the uncoupled channel formalism, then the coupled channel one,
and finally I explain the underlying attractive and repulsive eigenchannel
structure in the coupled triplets, which is a consequence of
the OPE tensor force.
This structure is what allows for the reduction of the number of subtractions
required to renormalize TPE in the coupled triplets from six to two.

\subsection{Uncoupled triplets}

Here I derive the formulas for the phase shifts in distorted wave perturbation
theory and then discuss how to renormalize them (in the sense of
removing the cutoff).

I begin by considering the reduced Schr\"odinger equations for two different
potentials
\begin{eqnarray}
  -u_k'' + \left[ 2\mu\,V_{\rm LO} + \frac{l(l+1)}{r^2} - k^2 \right]\,u_k(r)
  &=& 0
  \, , \\
  -v_k'' + \left[ 2\mu\,V_{\rm EFT} + \frac{l(l+1)}{r^2} - k^2 \right]\,v_k(r)
  &=& 0
  \, , 
\end{eqnarray}
where $V_{\rm LO}$ and $V_{\rm EFT}$ are the ${\rm LO}$ and full (i.e. computed
to all orders in the EFT expansion) potentials,  $u_k$, $v_k$ the respective
reduced wave functions, $\mu$ the reduced mass of the system,
$l$ the orbital angular momentum and $k$
the center-of-mass momentum.
From these two equations it is possible to build
the following Wronskian identity
\begin{eqnarray}
  - (u_k' v_k - u_k v_k')\Big|_{R_a}^{R_b} =
  \int_{R_a}^{R_b} (V_{\rm EFT} - V_{\rm LO}) v_k(r) u_k(r)\,dr \, , \nonumber \\
\end{eqnarray}
with $R_a$ and $R_b$ the lower and upper integration radii.
If one chooses the asymptotic normalization
\begin{eqnarray}
  u_k(r) &\to& k^{l}\, \left[
    \cot{\delta_{\rm LO}} \hat{j}_l(k r) - \hat{y}_l(k r)
    \right] \, , \\
  v_k(r) &\to& k^{l}\, \left[
    \cot{\delta_{\rm EFT}} \hat{j}_l(k r) - \hat{y}_l(k r)
    \right] \, ,
\end{eqnarray}
where $\hat{j}_l(x) = x\,j_l(x)$, $\hat{y}_l(x) = x\,y_l(x)$
(with $j_l(x)$ and $y_l(x)$ the spherical Bessel functions) and
$\delta_{\rm LO}$, $\delta_{\rm EFT}$ are the ${\rm LO}$ and full phase shifts, 
and then takes the limits $R_a \to 0$ and $R_b \to \infty$,
one ends up with the expression
\begin{eqnarray}
  && \cot{\delta_{\rm EFT}} - \cot{\delta_{\rm LO}} = \nonumber \\
  && \quad \frac{2\mu}{k^{2l+1}}\,
  \int_0^{\infty}\,dr\,u_k(r)\,(V_{\rm EFT} - V_{\rm LO})\,v_k(r)
  \, ,
\end{eqnarray}
which is exact for a regular or a regularized potential~\footnote{In
    principle the right hand side of the equation contains a Wronskian
    term of the type $(u_k v_k' - u_k' v_k)$ and evaluated at the origin
    (or at a boundary radius if one uses boundary condition
    regularization). This term is identically zero if
    the potential is regular or has been regularized. It if has
    not been regularized, then it behaves as a subtraction
    constant at the origin (or boundary radius). Its inclusion
    is redundant though after the regularization of the integral
    in the right hand side, which is why it can be safely ignored.
    A brief discussion can be found in Appendix A of
    Ref.~\cite{Valderrama:2009ei}.}.
If one now explicitly considers the EFT series for the full quantities
\begin{eqnarray}
 && V_{\rm EFT} \to V_{\rm LO} + \sum_{\nu > \nu_{\rm LO}} V^{(\nu)} \, , \\ 
 && v_k(r) \to u_k(r) + \sum_{\nu > \nu_{\rm LO}} u_k^{(\nu)} \, , \\
 &&
  \cot{\delta_{\rm EFT}} \to \cot{\delta_{\rm LO}}
  + \sum_{\nu > \nu_{\rm LO}} [\cot{\delta}]^{(\nu)} \, , 
\end{eqnarray}
and expands according to power counting
(with $\nu_{\rm LO}$ the order at which the ${\rm LO}$ potential
enters --- $\nu_{\rm LO} = -1$ in my convention --- and where
I ignore for the moment the iteration of subleading terms,
which does not happen till $\nu = 5$), one arrives to
\begin{eqnarray}
  - \frac{\delta^{(\nu)}}{\sin^2 \delta_{\rm LO}} = \frac{2\mu}{k^{2l+1}}\,
  \int_0^{\infty}\,dr\,V^{(\nu)}(r)\,u_k^2(r)\, , \label{eq:delta-pert}
\end{eqnarray}
from which the perturbative correction to the phase shift
can be calculated~\footnote{
  The left- and right-hand-side of Eq.~(\ref{eq:delta-pert}) diverge
  for $\sin{\delta}_{\rm LO} \to 0$,
  yet this is not a  problem for the calculation of
  the perturbative phase shift.
  This can be appreciated by simply multiplying both sides by
  ${\sin}^2\,{\delta}_{\rm LO}$, which is equivalent to changing
  the asymptotic normalization of the wave function to
  $u_k(r) \to k^l\,[\cos{\delta_{\rm LO}} \hat{j}_l(k r) -
    \sin{\delta_{\rm LO}}\,\hat{y}_l(k r)]$.
}.

At this point it is in order to briefly explain the power counting
conventions I am using:
\begin{itemize}
\item[(i)] For the EFT potential, its power counting index refers to its
  original scaling in momentum space. Strictly speaking, the configuration
  space potential is shifted by three orders, or $(Q/M)^3$, as
  a consequence of the integration over momenta
  from the Fourier-transform.
\item[(ii)] Each iterations contributes a factor of $(Q/M)$, which originates
  from integrating the two-nucleon loops in momentum space. This implies that
  only the $\nu = -1$ piece of the EFT potential is iterated.
\item[(iii)] The first few terms in the EFT two-nucleon potential for
  the triplets are
  \begin{eqnarray}
    V_{\rm EFT} &=& 
    \underbrace{V_{\rm LO} (=V^{(-1)})}_{\rm OPE} +
    \underbrace{V^{(2)}}_{\rm TPE(L)} +
    \underbrace{V^{(3)}}_{\rm TPE(SL)} \nonumber \\ &+& \dots \, , 
  \end{eqnarray}
  where ${\rm LO}$ is taken to be $\nu = -1$ (because of the convention
  I use for counting loops) and contains OPE plus the necessary
  contacts to renormalize, while $\nu = 2$, $3$ are comprised of
  leading and subleading TPE and the relevant contacts.
  This is the reason why second order distorted wave perturbation
  theory does not appear till $\nu =5$ (unless one decides to
  include contacts at $\nu = 0$, $1$).
\item[(iv)] For simplicity, the power counting index of all EFT quantities
  (with the exception of the momentum space potential) will be {\it nominal}
  instead of {\it actual}. That is, the index refers to the contributions
  of the potential that are included in the calculation of said quantity.
\end{itemize}
Other works use different conventions, which might cause confusion if not
properly noticed. For instance, in~\cite{Long:2011qx,Long:2011xw}
the contributions to the EFT potential are labeled as in NDA (and
in addition the TPE potential is enhanced by one order
with respect to here).

Returning to the analysis of distorted wave perturbation theory,
if one is interested in renormalizability there is the problem that the
integral appearing in Eq.~(\ref{eq:delta-pert}) does not have
a clear subtraction structure: the reason is the existence of
an energy-dependent normalization factor in front of $u_k$,
which ruins naive attempts to make direct subtractions.
Basically the $k^2$ expansion of the wave function is given by
\begin{eqnarray}
  u_k(r) = \mathcal{A}(k)\,\hat{u}_k(r) \, ,
\end{eqnarray}
with
\begin{eqnarray}
  \hat{u}_k(r) = \sum_{n=0}^{\infty} k^{2n} \hat{u}_{2n}(r) \, ,
\end{eqnarray}
where $\hat{u}_{2n}$ are increasingly suppressed as $r \to 0$, with
\begin{eqnarray}
  \hat{u}_{2n}(r) \propto r^{\frac{3}{4} + \frac{5 n}{2}}\,
  f_{2n}(2 \sqrt{\frac{a_3}{r}}) \, ,
\end{eqnarray}
with $f_{2n}(x)$ a set of functions that depend on the sign of the tensor force.
The length scale $a_3$ measures the strength of the OPE potential in the
triplet channels at distances below the pion Compton wavelength, i.e.
\begin{eqnarray}
  \lim_{m r \to 0}\, 2\mu\,V_{\rm OPE}(r) \to \pm \frac{a_3}{r^3} \, ,
\end{eqnarray}
where $m$ refers to the pion mass.
For attractive triplets one has a combination of a sine and a cosine
\begin{eqnarray}
  f_{2n}(x) \sim a_s\,\sin(x) + a_c\,\cos(x) \, ,
\end{eqnarray}
while for repulsive triplets one has a decreasing exponential
\begin{eqnarray}
  f_{2n}(x) \sim b\,e^{-x} \, ,
\end{eqnarray}
where $a_s$, $a_c$ and $b$ are numerical constants.
Notice that there is the possibility of adding a growing exponential solution
in the repulsive case: however, this is not physically acceptable as it
generates a wave function that is not only irregular at short
distances but indeed has an essential singularity
at $r \to 0$.

From the previous it is evident why a naive subtraction scheme does not work:
the $k^2$ expansion of $u_k(r)$ does not generate terms that are more and
more suppressed at shorter distances.
Yet, by removing an energy dependent normalization factor (which bears some
resemblance to the Jost function of non-singular potentials),
the subsequent $k^2$ expansion has good properties.

Once the momentum-dependent normalization factor is considered,
the perturbative phase shift reads
\begin{eqnarray}
  - \frac{\delta^{(\nu)}}{\sin^2 \delta_{\rm LO}} = \frac{2\mu}{k^{2l+1}}\,
  \mathcal{A}^2(k)\,\int_0^{\infty}\,dr\,V^{(\nu)}(r)\,\hat{u}_k^2(r)\, .
  \nonumber \\
\end{eqnarray}
If one takes into account that the EFT potential can be decomposed into a
finite- and contact-range part
\begin{eqnarray}
  V^{(\nu)} = V_F^{(\nu)} + V_C^{(\nu)} \, ,
\end{eqnarray}
regularizes the finite-range part of the EFT potential as
\begin{eqnarray}
  V_F^{(\nu)}(r; R_c) = V_F^{(\nu)}(r)\,\theta(r - R_c) \, ,
\end{eqnarray}
and momentarily ignores the contact-range part, then regularizing the
perturbative phase shifts reduces to the problem of making the
following integral finite:
\begin{eqnarray}
  \hat{I}_k^{(\nu)}(R_c) = \int_{R_c}^{\infty}\,dr\,V_F^{(\nu)}(r)\,\hat{u}_k^2(r)\, .
\end{eqnarray}
By taking into account the $r \to 0$ behavior of the $\nu$-th order
contribution to the finite-range EFT potential:
\begin{eqnarray}
  V_F^{(\nu)}(r) \propto \frac{1}{r^{3+\nu}} \, ,
\end{eqnarray}
and the $k^2$ expansion of $\hat{u}_k^2$, it becomes trivial to determine
the superficial degree of divergence of the previous integral 
\begin{eqnarray}
  \hat{I}_k^{(\nu)}(R_c) =
  \sum_{n_1, n_2}  k^{2(n_1 + n_2)} \int_{R_c}^{\infty}\,dr\,V_F^{(\nu)}(r)\,
  \hat{u}_{2 n_1}(r)\,\hat{u}_{2 n_2}(r) && \nonumber \\
  \propto \sum_{n=0}^{\infty} k^{2n} \int_{R_c}
  \,\frac{dr}{r^{3/2+\nu - 5/2\,n}} \, , && \nonumber \\
\end{eqnarray}
where in the second line only the lower integration bound is shown
to emphasize the divergences (indeed, the expression
in the second line is only valid for $m R_c \ll 1$,
with $m$ the pion mass).

Once one takes into account that the wave functions contain a trigonometric
or exponential component, depending on the sign of the tensor force,
it is apparent that for the attractive case the perturbative
integral is divergent if the following condition is met
\begin{eqnarray}
  \frac{3}{2} + \nu - \frac{5 n}{2} \geq 1 \, .
\end{eqnarray}
Thus, subleading and leading TPE ($\nu = 2, 3$) are perturbatively
renormalizable with two subtractions of the perturbative integral:
\begin{eqnarray}
  \hat{I}_k^{(\nu)}(R_c) = 
  \int_0^{\infty}\,dr\,V_F^{(\nu)}(r)\,\hat{u}_k^2(r) + \lambda_0 + \lambda_2 k^2
  \, , \label{eq:Ik-two-subs}
\end{eqnarray}
which now has a well-defined $R_c \to 0$ limit.
Alternatively, one might use instead a regularized contact-range potential
with two couplings: provided a good choice of
regulator~\cite{PavonValderrama:2025tmp} (i.e. a regulator that does
not generate exceptional cutoffs or other pathologies),
it will be equivalent to the previous subtractions. 
A really simple choice is an energy-dependent, local delta-shell regulator:
\begin{eqnarray}
  V_C(r; R_c) = \sum_{n} C_{2n}(R_c) k^{2n}\,\frac{\delta(r - R_c)}{4 \pi R_c^2} \, ,
\end{eqnarray}
where the relation between the subtraction parameters $\lambda_{2n}$
and the couplings $C_{2n}$ is linear.

In contrast, for the repulsive case the reduced wave function is exponentially
suppressed at short distances, and the $k^2$ expansion of the perturbative
integral will only contain non-divergent terms:
\begin{eqnarray}
  \int_{R_c}
  \,dr\,\frac{e^{-4 \sqrt{\frac{a_3}{r}}}}{r^{3/2+\nu - 5/2\,n}}
  \quad \mbox{(convergent)} \, .
\end{eqnarray}
That is, the perturbative renormalizability of two-pion exchange
in a repulsive triplet does not need the inclusion of counterterms.

\subsection{Coupled triplets}

For the coupled channels one begins by writing the reduced Sch\"odinger
equations in matrix form
\begin{eqnarray}
  -{\bf u}_k'' + \left[ 2\mu\,{\bf V}_{\rm LO} + \frac{{\bf L}^2}{r^2} - k^2 \right]\,{\bf u}_k &=& 0 \, ,
\end{eqnarray}
where for concreteness I have only considered the ${\rm LO}$ case and with
the potential and reduced wave function now being $N \times N$ matrices.
The angular momentum operator is a diagonal matrix given by
\begin{eqnarray}
  {\bf L}^2 = {\rm diag}\left( l_1(l_1 + 1), \dots, l_N(l_N+1) \right) \, ,
\end{eqnarray}
with $l_i$ the orbital angular momentum of channel $1 \leq i \leq N$.
The asymptotic form of the reduced wave function is
\begin{eqnarray}
  {\bf u}_k(r) \to {\bf J}_k(r)\,{\bf M}_{\rm LO}(k) - {\bf Y}_k(r) \, ,
\end{eqnarray}
with ${\bf M}_{\rm LO}(k)$ the matrix equivalent of $\cot{\delta_{\rm LO}}$
in the uncoupled channel case and ${\bf J}_k$ and ${\bf Y}_k$
diagonal matrices defined as
\begin{eqnarray}
  {\bf J}_k(r) &=& {\rm diag}\,
  \big( \hat{j}_{l_1}(kr), \dots, \hat{j}_{l_N}(k r) \big) \, , \\
  {\bf Y}_k(r) &=& {\rm diag}\,
  \big( \hat{y}_{l_1}(kr), \dots, \hat{y}_{l_N}(k r) \big) \, . 
\end{eqnarray}
By considering the corresponding reduced Schr\"odinger equation for the
full expansion of the EFT potential and constructing a suitable
Wronskian identity, one arrives at the expression
\begin{eqnarray}
  \delta\,{\bf M}(k) &=& {\bf M}_{\rm EFT}(k) - {\bf M}_{\rm LO}(k)
  \nonumber \\
  &=& \frac{2\mu}{k}\,\int_0^{\infty}\,dr\,{\bf v}_k^T\,({\bf V}_{\rm EFT} - {\bf V}_{\rm LO})\,{\bf u}_k \, ,
\end{eqnarray}
which is exact for regular or regularized potentials (while still
being formally correct with non-regularized singular potentials),
where ${\bf M}_{\rm EFT}$, ${\bf V}_{\rm EFT}$ and ${\bf v}_k$ refer to
quantities calculated at all orders in the EFT expansion.
If one expands and ignores the iteration of subleading order
terms, the expression for the first order distorted wave
perturbations is
\begin{eqnarray}
  {\bf M}^{(\nu)}(k) 
  &=& \frac{2\mu}{k}\,\int_0^{\infty}\,dr\,
  {\bf u}_k^T\,{\bf V}^{(\nu)}\,{\bf u}_k \, ,
\end{eqnarray}
which is valid for $\nu < 5$ within the power counting conventions I am using.

For rendering the previous formula into a more familiar form, one begins
by taking into account that for coupled triplets $N=2$ with the 
${\bf M}$ matrix given by
\begin{eqnarray}
  {\bf M}(k) = {\bf R}(\epsilon)\,
  \begin{pmatrix}
    \cot{\delta_{\alpha}} & 0 \\
    0 & \cot{\delta_{\beta}}
  \end{pmatrix}\,{\bf R}(-\epsilon)\,
  \label{eq:M-def}
\end{eqnarray}
where $\delta_{\alpha}$ and $\delta_{\beta}$ are the eigen phase
shifts~\cite{PhysRev.86.399},
and ${\bf R}$ a rotation matrix:
\begin{eqnarray}
  {\bf R}(\epsilon) =
    \begin{pmatrix}
    \cos{\epsilon} & -\sin{\epsilon} \\
    \sin{\epsilon} & \cos{\epsilon}
    \end{pmatrix} \, ,
      \label{eq:R-def}
\end{eqnarray}
with $\epsilon$ the mixing angle.
Within this parametrization the ${\bf u}_k {\bf R}$ product is given by
\begin{eqnarray}
  {\bf u}_k(r) {\bf R}(\epsilon) =
  \begin{pmatrix}
    u_{k\alpha}(r) & u_{k\beta}(r) \\
    w_{k\alpha}(r) & w_{k\beta}(r) 
  \end{pmatrix}
  \,
  \begin{pmatrix}
    \frac{1}{k^{l_a}} & 0 \\
    0 & \frac{1}{k^{l_b}}
  \end{pmatrix} \, , \,\,
\end{eqnarray}
where the $\alpha$ and $\beta$ scattering solutions behave asymptotically as
\begin{eqnarray}
  \begin{pmatrix}
    u_{k\alpha}(r) \\
    w_{k\alpha}(r) \\
  \end{pmatrix} \to
  \begin{pmatrix}
    \phantom{+}k^{l_a}\,\cos{\epsilon}\,(\cot{\delta_{\alpha}}\,\hat{j}_{l_a}(k r) -
    \hat{y}_{l_a}(k r)) \\
    \phantom{+}k^{l_a}\,\sin{\epsilon}\,(\cot{\delta_{\alpha}}\,\hat{j}_{l_b}(k r) -
    \hat{y}_{l_b}(k r))
  \end{pmatrix} \, , \nonumber \\ \\
    \begin{pmatrix}
    u_{k\beta}(r) \\
    w_{k\beta}(r) \\
  \end{pmatrix} \to
  \begin{pmatrix}
    -k^{l_b}\,\sin{\epsilon}\,(\cot{\delta_{\beta}}\,\hat{j}_{l_a}(k r) -
    \hat{y}_{l_a}(k r)) \\
    \phantom{+}k^{l_b}\,\cos{\epsilon}\,(\cot{\delta_{\beta}}\,\hat{j}_{l_b}(k r) -
    \hat{y}_{l_b}(k r))
  \end{pmatrix} \, , \nonumber \\
\end{eqnarray}
with $l_a = j-1$ and $l_b = j+1$, and $j$ being
the total angular momentum of
the coupled channel.

If one now expands in terms of the components of the matrices (and in terms
of the EFT expansion, assuming $\nu < 5$), one arrives at
\begin{eqnarray}
  {[\cot{\delta_{\alpha}}]}^{(\nu)} =
  - \frac{\delta_{\alpha}^{(\nu)}}{\sin^2{\delta_{\alpha {\rm LO}}}} &=&
  \frac{2\mu}{k^{2j-1}}\,I^{(\nu)}_{\alpha \alpha}(k) \, , \\
  {[\cot{\delta_{\beta}}]}^{(\nu)} =
  - \frac{\delta_{\beta}^{(\nu)}}{\sin^2{\delta_{\beta {\rm LO}}}}
  &=& \frac{2\mu}{k^{2j+3}}\,I^{(\nu)}_{\beta \beta}(k) \, , \\
  \epsilon^{(\nu)}\,\left(
  \cot{\delta_{\alpha {\rm LO}}} - \cot{\delta_{\beta {\rm LO}}} \right) &=&
  \frac{2\mu}{k^{2j+1}}\,I^{(\nu)}_{\alpha \beta}(k) \, ,
\end{eqnarray}
with the individual perturbative integrals given by
\begin{eqnarray}
  I_{\rho \sigma}^{(\nu)}(k) &=& \int_0^{\infty}\,dr\,\Big[
    u_{k\rho}(r) V_{aa}^{(\nu)}(r) u_{k\sigma}(r) \nonumber \\
    && + V_{ab}^{(\nu)}(r) \left(
    u_{k\rho}(r) w_{k\sigma}(r) +
    w_{k\rho}(r) u_{k\sigma}(r) 
    \right) \nonumber \\
    && + w_{k\rho}(r)\, V_{bb}^{(\nu)}(r)\,w_{k\sigma}(r) \Big] \, ,
  \label{eq:perturbative-integrals-eigen}
\end{eqnarray}
where $\rho, \sigma \in \{ \alpha, \beta \}$ depending on the case and with
$V_{cd}^{(\nu)}$ the components of 
\begin{eqnarray}
  {\bf V}^{(\nu)}(r) =
  \begin{pmatrix}
    V^{(\nu)}_{aa}(r) & V^{(\nu)}_{ab}(r) \\
    V^{(\nu)}_{ba}(r) & V^{(\nu)}_{bb}(r)
  \end{pmatrix} \, ,
\end{eqnarray}
which are symmetrical ($V_{ab}^{(\nu)} = V_{ba}^{(\nu)}$)
in the case of the configuration space potential.

Once the perturbative corrections for the eigen phase shifts have been
calculated, one might transform them into the nuclear bar
parametrization~\cite{PhysRev.105.302} by expanding
the well-known conversion formulas
\begin{eqnarray}
  \delta_{\alpha} + \delta_{\beta} &=& \bar{\delta}_{\alpha} + \bar{\delta}_{\beta}
  \, , \\
  \sin{(\delta_{\alpha} - \delta_{\beta})} &=&
  \frac{\sin{2 \bar{\epsilon}}}{\sin{2 \epsilon}}  \, , \\
  \sin{(\bar{\delta}_{\alpha} - \bar{\delta}_{\beta})} &=&
  \frac{\tan{2 \bar{\epsilon}}}{\tan{2 \epsilon}} \, , 
\end{eqnarray}
in terms of the power counting. Even though their expansion is
straightforward and poses no complication, the expressions
thus obtained are rather involved and will not be
written down here.

For analyzing the divergence pattern of the perturbative integral, one
follows the same steps as in the uncoupled channel case, beginning
with the $k^2$ expansion of the reduced wave function.
This again requires to explicitly take into account the existence of
an energy-dependent normalization factor:
\begin{eqnarray}
  \begin{pmatrix}
    u_{k \rho}(r) \\
    w_{k \rho}(r) 
  \end{pmatrix} =
  \mathcal{A}_{\rho}(k)\, 
 \begin{pmatrix}
    \hat{u}_{k \rho}(r) \\
    \hat{w}_{k \rho}(r) 
 \end{pmatrix} \, ,
 \label{eq:hat-normalization}
\end{eqnarray}
where $\rho \in \{ \alpha, \beta \}$
indicates the linearly independent solution.
In the new normalization, the $k^2$ expansion of the reduced wave function
is
\begin{eqnarray}
 \begin{pmatrix}
    \hat{u}_{k \rho}(r) \\
    \hat{w}_{k \rho}(r) 
 \end{pmatrix} =
 \sum_{n=0}^{\infty}
 k^{2n} \, 
 \begin{pmatrix}
    \hat{u}_{2n \rho}(r) \\
    \hat{w}_{2n \rho}(r) 
  \end{pmatrix}
 \, ,
\end{eqnarray}
where each term is more suppressed than the previous one at short enough
distances:
\begin{eqnarray}
   \begin{pmatrix}
    \hat{u}_{2n \rho}(r) \\
    \hat{w}_{2n \rho}(r) 
   \end{pmatrix}
   \to
   r^{\frac{3}{4} + \frac{5 n}{2}}\,
      \begin{pmatrix}
    {f}_{2n \rho}(2\sqrt{\frac{a_{3A}}{r}},2\sqrt{\frac{a_{3R}}{r}}) \\
    {g}_{2n \rho}(2\sqrt{\frac{a_{3A}}{r}},2\sqrt{\frac{a_{3R}}{r}}) 
   \end{pmatrix} \, . \nonumber \\
\end{eqnarray}
Here, $a_{3A}$ and $a_{3R}$ are two lengths scales characterizing the strength of
the ${\rm LO}$ potential when $m r \to 0$, and $f_{2n \rho}$ and $g_{2n \rho}$
a linear combination of sine, cosine and a decaying exponential function
\begin{eqnarray}
  && f_{2n \rho}(x_A,x_R)\, , \, g_{2n \rho}(x_A,x_R) \nonumber \\
  && \qquad \sim a_s\,\sin{x_A} + a_c\,\cos{x_A}
  + b\,e^{- x_R} \, ,\nonumber \\
\end{eqnarray}
with $a_s$, $a_c$, $b$ real numbers determining
the specific linear combination.

From this the perturbative integral defined
in Eq.~(\ref{eq:perturbative-integrals-eigen}) can be rewritten as
\begin{eqnarray}
  I_{\rho \sigma}^{(\nu)}(k; R_c) &=& \mathcal{A}_{\rho}(k) \mathcal{A}_{\sigma}(k)\,
  \hat{I}_{\rho \sigma}^{(\nu)}(k; R_c) \, , 
\end{eqnarray}
where $\hat{I}_{\rho \sigma}^{(\nu)}$ has the same divergence structure
as its uncoupled channel counterpart for an attractive
triplet.
That is, by making two subtractions for the perturbative integral of each
of the $\alpha \alpha$, $\alpha \beta$ and $\beta \beta$ two channel
combinations, they will become finite when the cutoff is removed.

Thus, six subtractions (or counterterms if one uses a contact-range potential)
guarantee the perturbative renormalizability of leading and
subleading TPE, as originally pointed out
in~\cite{Valderrama:2009ei,Valderrama:2011mv}.
The question is whether this number can be further reduced.

\subsection{Attractive and repulsive eigenchannels}

The number of subtractions can indeed be reduced by taking into account
the existence of simplifications entailed by the structure of
the ${\rm LO}$ potential.
The OPE potential is composed of a spin-spin and tensor component
\begin{eqnarray}
  V_{\rm LO} = V_S\,\vec{\sigma}_1 \cdot \vec{\sigma}_2 +
  V_T\,(3\,\vec{\sigma}_1 \cdot \hat{r} \vec{\sigma}_2 \cdot \hat{r} -
  \vec{\sigma}_1 \cdot \vec{\sigma}_2) \, ,
\end{eqnarray}
with $\vec{\sigma}_{i}$ the Pauli matrices as applied to nucleon $i = 1,2$
and where the isospin dependence has not been explicitly written.
For a coupled channel with total angular momentum $j$, the matrix elements
of the spin-spin and tensor operators
in the corresponding spectroscopic basis (i.e. ${}^3S_1$-${}^3D_1$ for $j=1$,
${}^3P_2$-${}^3F_2$ for $j=2$ and so on) are
\begin{eqnarray}
  {\bf V}_{\rm LO} &=& V_S\,
  \begin{pmatrix}
    1 & 0 \\
    0 & 1 
  \end{pmatrix} \nonumber \\ &+&
  V_T\,
  \frac{1}{2j+1}\,
  \begin{pmatrix}
    -{2\,(j-1)} & {6\sqrt{j(j+1)}} \\
    {6\sqrt{j(j+1)}} & -{2\,(j+1)} 
  \end{pmatrix} \, . \nonumber \\
\end{eqnarray}
The ${\rm LO}$ potential can be diagonalized with the rotation
\begin{eqnarray}
  {\bf R}_j = \frac{1}{\sqrt{2j+1}}
  \begin{pmatrix}
    \sqrt{j+1} & \sqrt{j} \\
    -\sqrt{j} & \sqrt{j+1}
  \end{pmatrix}
\end{eqnarray}
which shows the existence of one attractive and one repulsive eigenvalue
for the tensor component of the potential:
\begin{eqnarray}
  {\bf R}_j {\bf V}_{\rm LO} {\bf R}_j^T =
  V_S\,
  \begin{pmatrix}
    1 & 0 \\
    0 & 1 
  \end{pmatrix} +
  V_T\,
  \begin{pmatrix}
    2 & 0 \\
    0 & -4 
  \end{pmatrix} \, .
\end{eqnarray}
Even though the previous transformation diagonalizes the OPE potential,
it does not diagonalize the orbital angular momentum operator, i.e.
the centrifugal barrier.
Yet, if one takes into account that the OPE potential diverges as $1/r^3$
at distances sufficiently small in comparison with the Compton wavelength
of the pion, it is apparent that it will overtake the centrifugal
barrier:
\begin{eqnarray}
    {\bf R}_j \left[ 2\mu\,{\bf V}_{\rm LO} + \frac{{\bf L}^2}{r^2}\right] {\bf R}_j^T \to \pm \frac{a_3}{r^3}\,  \begin{pmatrix}
    2 & 0 \\
    0 & -4 
  \end{pmatrix} \, ,
\end{eqnarray}
when $r \to 0$.
That is, there is an attractive and repulsive eigenchannel
at short enough distances.

If one writes the reduced Schr\"odinger equation in the diagonal basis as
\begin{eqnarray}
  -{\bf v}_k'' + {\bf R}_j\,\left[ 2\mu\,{\bf V}_{\rm LO} + \frac{{\bf L}^2}{r^2} - k^2 \right]\,{\bf R}_j^T\,{\bf v}_k &=& 0 \, ,
\end{eqnarray}
it is possible to choose the two linearly independent reduced wave functions
as a pure attractive or repulsive component at short enough distances
(by which it is meant when the strength of tensor OPE overcomes
the centrifugal barrier)
\begin{eqnarray}
  {\bf v}_k(r) &\to&
  \begin{pmatrix}
    v_A(r) & 0 \\
    0 & v_R(r)
  \end{pmatrix} \,\, \mbox{or} \,\, 
  \begin{pmatrix}
    v_R(r) & 0 \\
    0 & v_A(r)
  \end{pmatrix} \, ,
  \label{eq:vk-short}
\end{eqnarray}
for $r \to 0$, depending on which eigenchannel is attractive or repulsive,
i.e. on which of the following situations one has
\begin{eqnarray}
  && 2\mu\,{\bf R}_j {\bf V}_{\rm LO} {\bf R}_j^T \nonumber \\
  && \quad \to
  \frac{1}{r^3}\,  \begin{pmatrix}
    -a_{3A} & 0 \\
    0 & +a_{3R} 
  \end{pmatrix}
  \,\,\,\mbox{or}\,\,\,
    \frac{1}{r^3}\,  \begin{pmatrix}
    +a_{3R} & 0 \\
    0 & -a_{3A} 
    \end{pmatrix} \, , \nonumber \\
    \label{eq:VLO-short}
\end{eqnarray}
where $a_{3A}$ and $a_{3R}$ are the length scales characterizing the attractive
and repulsive eigenchannels~\footnote{Once the $a_{3A}$ and $a_{3R}$ scales
  are defined, it becomes apparent that Eqs.~(\ref{eq:vk-short}) and
  (\ref{eq:VLO-short}) are valid for $r \ll a_{3A}, a_{3R}$.
} .

The bottom-line is that the $\alpha$ and $\beta$ scattering states are linear
combinations of wave functions that at short distances behave as
purely attractive ($A$) or repulsive ($R$) solutions:
\begin{eqnarray}
  \begin{pmatrix}
    \hat{u}_{k\rho} \\
    \hat{w}_{k\rho}
  \end{pmatrix} =
  c_A^{\rho}\,
  \begin{pmatrix}
    \hat{u}_{A} \\
    \hat{w}_{A}
  \end{pmatrix}
  + c_R^{\rho}\,
  \begin{pmatrix}
    \hat{u}_{R} \\
    \hat{w}_{R}
  \end{pmatrix} \, ,
\end{eqnarray}
where $\rho \in \{ \alpha$, $\beta \}$ and $c_A^{\rho}$, $c_R^{\rho}$
the coefficients of this linear transformation.
This implies that the $\hat{I}_{\rho \sigma}^{(\nu)}(k)$ perturbative integrals can
be expressed as linear combinations of perturbative integrals
for the $A$ and $R$ wave functions defined above
\begin{eqnarray}
  \hat{I}_{\rho \sigma}^{(\nu)}(k) &=&
  c_A^{\rho} c_A^{\sigma}\, \hat{I}_{AA}^{(\nu)}(k) \nonumber \\
  &+&
  \left( c_A^{\rho} c_R^{\sigma} + c_R^{\rho} c_A^{\sigma} \right)\,
  \hat{I}_{AR}^{(\nu)}(k) \nonumber \\
  &+& c_R^{\rho} c_R^{\sigma}\,\hat{I}_{RR}^{(\nu)}(k) \, ,
\end{eqnarray}
where the new integrals are given by
\begin{eqnarray}
  \hat{I}_{AA}^{(\nu)}(k) &=& \int_0^{\infty}\,dr\,\Big[
    \hat{u}_{A}(r) V_{aa}^{(\nu)}(r) \hat{u}_{A}(r) \nonumber \\
    && + 2\,V_{ab}^{(\nu)}(r)\, 
    \hat{u}_{A}(r) \hat{w}_{A}(r) 
    + \hat{w}_{A}(r)\, V_{bb}^{(\nu)}(r)\,\hat{w}_{A}(r) \Big] \, , \nonumber \\
  \\
  \hat{I}_{AR}^{(\nu)}(k) &=& \int_0^{\infty}\,dr\,\Big[
    \hat{u}_{A}(r) V_{aa}^{(\nu)}(r) \hat{u}_{R}(r) \nonumber \\
    && + V_{ab}^{(\nu)}(r) \left(
    \hat{u}_{A}(r) \hat{w}_{R}(r) +
    \hat{w}_{R}(r) \hat{u}_{A}(r) 
    \right) \nonumber \\
    && + \hat{w}_{A}(r)\, V_{bb}^{(\nu)}(r)\,\hat{w}_{R}(r) \Big]
  \, ,\nonumber \\ \\
  \hat{I}_{RR}^{(\nu)}(k) &=& \int_0^{\infty}\,dr\,\Big[
    \hat{u}_{R}(r) V_{aa}^{(\nu)}(r) \hat{u}_{R}(r) \nonumber \\
    && + 2\,V_{ab}^{(\nu)}(r)\, 
    \hat{u}_{R}(r) \hat{w}_{R}(r) 
    + \hat{w}_{R}(r)\, V_{bb}^{(\nu)}(r)\,\hat{w}_{R}(r) \Big] \, . \nonumber \\
\end{eqnarray}
Owing to the short-range behavior of the $A$ and $R$ wave functions,
which is analogous to that of uncoupled attractive and repulsive
wave functions, and following the previous arguments
for the uncoupled channel case,
it turns out that only one of these integrals is divergent:
$\hat{I}_{AA}^{(\nu)}$.
The other two, $\hat{I}_{AR}^{(\nu)}$ and $\hat{I}_{RR}^{(\nu)}$,
are finite thanks to the exponential suppression of
the wave function for the repulsive triplets.
The ultraviolet behavior of their integrands is
\begin{eqnarray}
  \hat{I}_{AR}^{(\nu)} \sim \int_{R_c}\,dr\,\frac{e^{-2 \sqrt{\frac{a_{3R}}{r}}}}{r^{3/2+\nu - 5/2\,n}} \, , \\
  \hat{I}_{RR}^{(\nu)} \sim \int_{R_c}\,dr\,\frac{e^{-4 \sqrt{\frac{a_{3R}}{r}}}}{r^{3/2+\nu - 5/2\,n}} \, , 
\end{eqnarray}
which is indeed convergent for $R_c \to 0$.
That is, for coupled triplets only two subtractions are required to render
the perturbative integral finite when the cutoff is removed.

\section{Regularization and renormalization}
\label{sec:reg-and-ren}

In this Section it is explained how to do the subleading order calculations
of the phase shifts with two counterterms. The cutoff dependence will be
carefully analyzed too, and it will be shown that the calculations
do indeed converge smoothly with respect to the cutoff.

\subsection{Calculation of the phase shifts}

I first consider the regularization and renormalization of the ${\rm LO}$
wave functions by means of boundary conditions at the cutoff
(or boundary) radius $R_c$.

Regardless of the specific regularization chosen, the ${\rm LO}$ wave functions
are expected to fulfill the following two principles:
\begin{enumerate}
\item[(i)] The wave functions are regular for $r \to 0$, that is:
  \begin{eqnarray}
    \lim_{r \to 0^+} u_k(r) = \lim_{r \to 0^+} w_k(r) = 0 \, .
  \end{eqnarray}
\item[(ii)] Any two wave functions that are not identical are orthogonal
  \begin{eqnarray}
    \int_0^{\infty} \,dr\,\left[ u_{k,\rho}(r)\,u_{k',\sigma}(r) +
      w_{k,\rho}(r)\,w_{k',\sigma}(r) \right] = 0 \, ,
    \nonumber \\
  \end{eqnarray}
  for $(k,\sigma) \neq (k',\rho)$, where $\rho$, $\sigma$ are indices labeling
  the solutions (notice that they do not have to coincide with the $\alpha$,
  $\beta$ scattering states).
\end{enumerate}

If one is using a potential-type regulator for the contact-range interaction
(for instance, a Gaussian), these two conditions will be trivially met
(provided the finite-range part of the interaction
has been regularized too).

In contrast if one is representing the short-range physics by a boundary
condition, one will have to impose the previous two conditions
explicitly at the cutoff radius:
\begin{itemize}
\item[(i)] For guaranteeing a regular wave function in the $R_c \to 0$ limit,
  I choose the condition
  \begin{eqnarray}
    u(R_c) = f_j\,w(R_c) \, , \label{eq:uw-reg}
  \end{eqnarray}
  where $f_j = \sqrt{(j+1)/j}$ for $j$ odd and $-\sqrt{j/(j+1)}$ for $j$ even.
  Other regularity conditions are possible and a few have been
  explored in~\cite{PavonValderrama:2005gu}.
  The advantage of the previous choice is that it has a straightforward
  interpretation in terms of the eigenchannels of the OPE potential:
  $u(R_c) - f_j w(R_c)$ corresponds to the repulsive part of
  the wave function at the cutoff radius.
\item[(ii)] For the orthogonality condition, it can be recast as a Wronskian
  equation at the cutoff radius~\cite{PavonValderrama:2005gu}:
  \begin{eqnarray}
    W(u_{k, \rho}, u_{k', \sigma}){\Big|}_{R_c} +
    W(w_{k, \rho}, w_{k', \sigma}){\Big|}_{R_c} &=& 0 \, ,
    \label{eq:Wronskian-coupled}
  \end{eqnarray}
  where $W(f,g) = f g' - f' g$ is the Wronskian. By expanding the Wronskians
  and combining them with the regularity condition,
  one arrives at
  \begin{eqnarray}
    && \frac{w_{k, \rho}'(R_c) + f_j u_{k, \rho}'(R_c)}{u_{k, \rho}(R_c)} =
    \nonumber \\
    && \qquad \qquad
    \frac{w_{k', \sigma}'(R_c) + f_j u_{k', \sigma}'(R_c)}{u_{k', \sigma}(R_c)} \, ,
    \label{eq:orthogonality-mod}
  \end{eqnarray}
  which is more convenient than the original orthogonality condition.
\end{itemize}

The previous conditions are by themselves insufficient to determine
the scattering wave functions (or, equivalently, the phase shifts).
They have to be supplemented by a renormalization condition fixing
one observable.

If one chooses to fix the $^3S_1$ ($^3P_2$) scattering length (volume),
the starting point is the asymptotic form of the $\alpha$ scattering
state at zero energy (and total angular momentum $j$):
\begin{eqnarray}
  u_{0\alpha} &\to& \frac{(2j-3)!!}{r^{j-1}} - \frac{r^j}{(2j-1)!!}\,
  \frac{1}{\alpha_{1j}} \, , \\
  w_{0\alpha} &\to& \frac{(2j-1)!!}{r^{j+1}}\,\frac{e_j}{\alpha_{1j}} \, , 
\end{eqnarray}
with $\alpha_{1j}$ and $e_j$ the scattering hypervolumes for the $\delta_{\alpha}$
eigen phase shift and the $\epsilon$ mixing angle, respectively.
This apparently requires two data points, yet $e_j$ is actually
fully determined from the regularity condition at
the cutoff radius, i.e. $u_{0\alpha}(R_c) = f_j w_{0\alpha}(R_c)$.
From here and the modified orthogonality conditions of
Eq.~(\ref{eq:orthogonality-mod}) it is straightforward to calculate
the phase shifts at arbitrary momentum by constructing the $\alpha$
and $\beta$ scattering solutions.

My choices of scattering length and volume for the $^3S_1$ and $^3P_2$ partial
waves are $\alpha_{1j} = 5.419\,{\rm fm}$ and $-0.04\,{\rm fm^3}$,
which coincide with the values originally used
in~\cite{Valderrama:2009ei,Valderrama:2011mv}.

Once the ${\rm LO}$ phase shifts have been determined, for the regularization of
the subleading results I use a local contact-range potential of the type
\begin{eqnarray}
  {\bf V}_{C}(r; R_c) = \left[ {\bf C}_0(R_c) + {\bf C}_2(R_c)\,k^2 \right]\,
  \frac{\delta(r-R_c)}{4 \pi R_c^2} \, ,  \label{eq:delta-shell-coupled}
\end{eqnarray}
where I choose ${\bf C}_{2n}$ that only act on the lowest orbital angular
momentum channel:
\begin{eqnarray}
  {\bf C}_{2n} =
  \begin{pmatrix}
    C_{2n} & 0 \\
    0 & 0
  \end{pmatrix} \, , \label{eq:contacts-coupled}
\end{eqnarray}
which corresponds to an S- (P-) wave contact in the $^3S_1$ ($^3P_2$) case.
This contact-range potential is equivalent to the set of subtractions
\begin{eqnarray}
  \lambda_{2n,\rho \sigma} = C_{2n}(R_c)\,
  \frac{u_{k \rho}(R_c) u_{k \sigma}(R_c)}{4 \pi R_c^2} \, , \nonumber \\
  \label{eq:contacts-to-subtractions}
\end{eqnarray}
with $\rho, \sigma \in \{ \alpha, \beta \}$.

At this point it is interesting to notice that, for the specific boundary
condition choice of Eq.~(\ref{eq:uw-reg}), it does not matter
whether the contacts act on a different partial wave.
Indeed, the alternative structures 
\begin{eqnarray}
  {\bf C}_{2n} =
  \begin{pmatrix}
    0 & C_{2n} \\
    C_{2n} & 0
  \end{pmatrix} \quad \mbox{or} \quad
  \begin{pmatrix}
    0 & 0 \\
    0 & C_{2n}
  \end{pmatrix} \, , \,
\end{eqnarray}
generate the same results as Eq.~(\ref{eq:contacts-coupled}),
with the only difference being the values of the couplings
(that only change by a proportionality constant).
However, this will not be true for other choices of
the boundary condition.

Finally, for determining the couplings $C_0$ and $C_2$,
I fit the $^3S_1$ and $^3P_2$ (nuclear bar~\cite{PhysRev.105.302})
subleading phase shifts to the Nijmegen II ones~\cite{Stoks:1994wp}
(which provide phase shifts that are within the uncertainties of the
ones obtained from the partial wave analysis of Ref.~\cite{Stoks:1993tb})
at $k = 100$ and $200\,{\rm MeV}$.
This is different to what I originally did
in Refs.~\cite{Valderrama:2009ei,Valderrama:2011mv}, where the phase shifts
were fitted within a range, but it has the advantage of making the analysis
of the cutoff dependence more transparent.
This latter point is important in view of the recent interest on whether
distorted wave perturbative calculations are actually cutoff
independent~\cite{Gasparyan:2022isg,Peng:2024aiz,Peng:2025ykg,Yang:2024yqv,PavonValderrama:2025tmp}.

\begin{figure*}[ttt]
  \begin{center}
    \includegraphics[width=5.6cm]{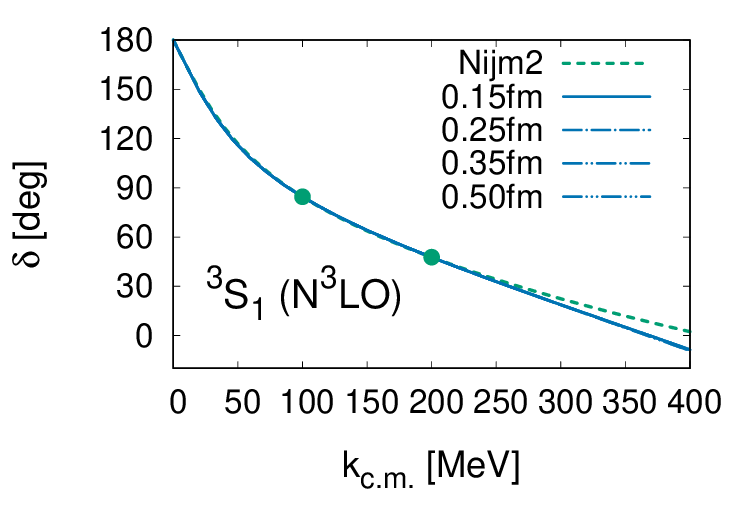}
    \includegraphics[width=5.6cm]{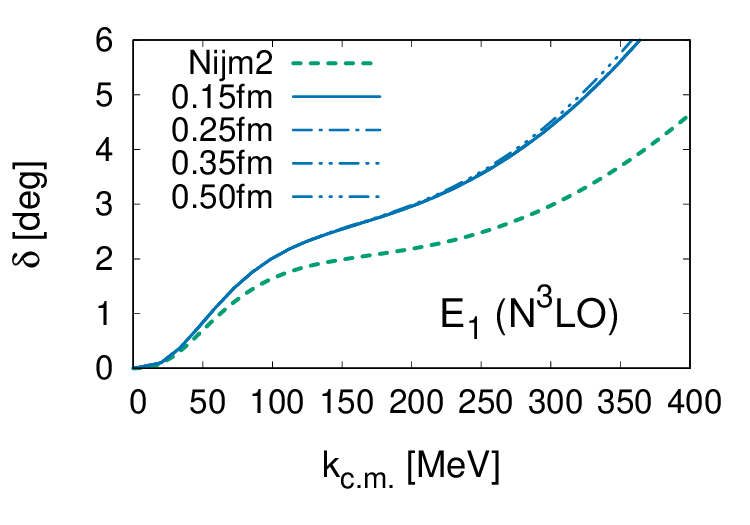}
    \includegraphics[width=5.6cm]{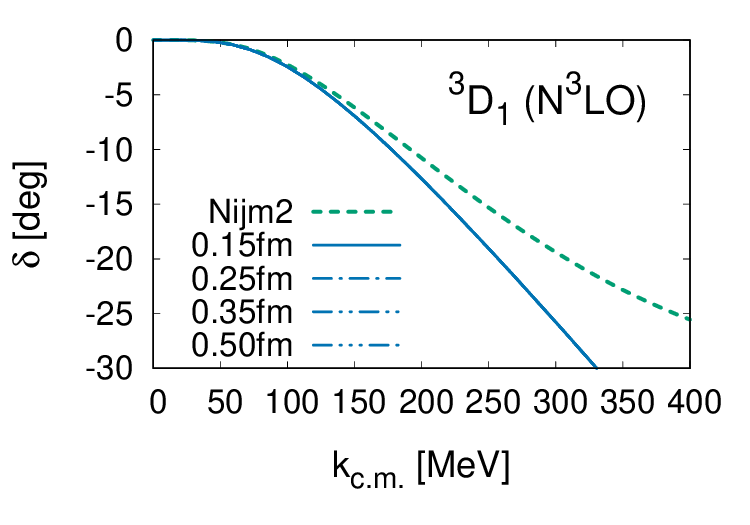}
    \includegraphics[width=5.6cm]{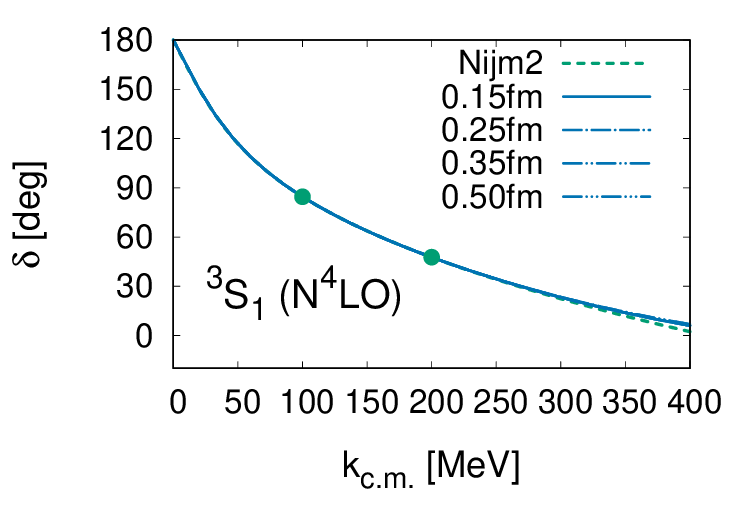}
    \includegraphics[width=5.6cm]{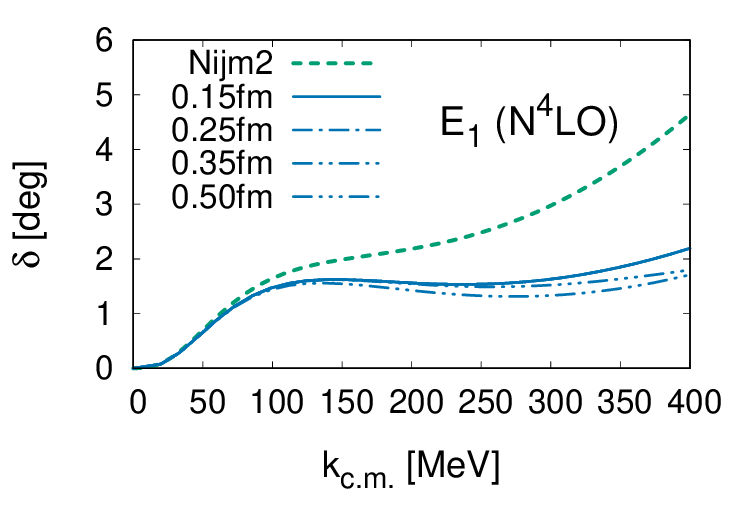}
    \includegraphics[width=5.6cm]{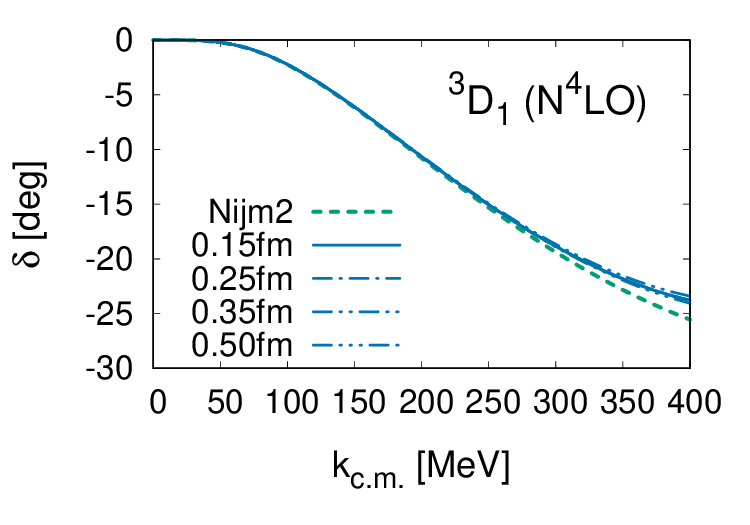}
    \includegraphics[width=5.6cm]{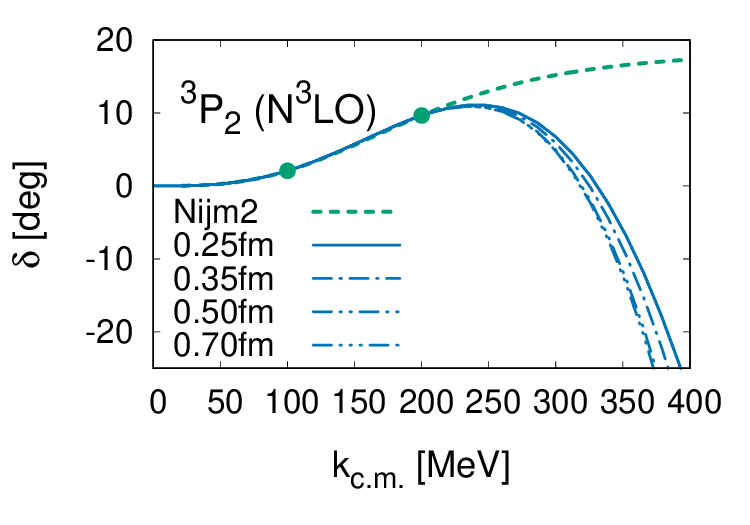}
    \includegraphics[width=5.6cm]{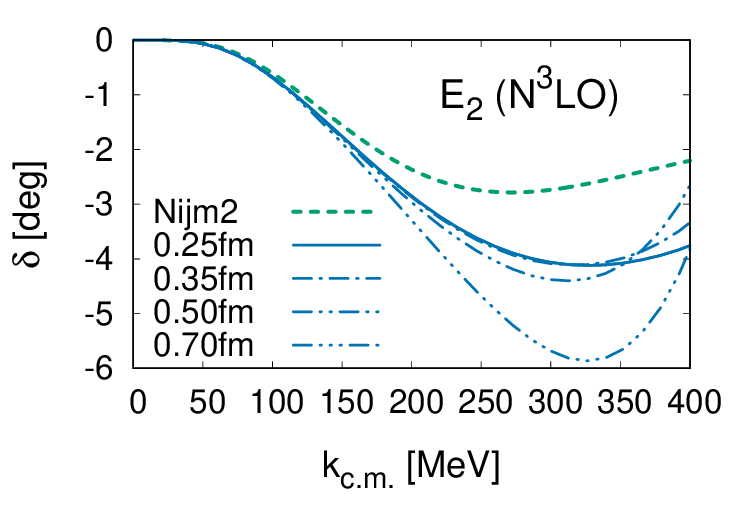}
    \includegraphics[width=5.6cm]{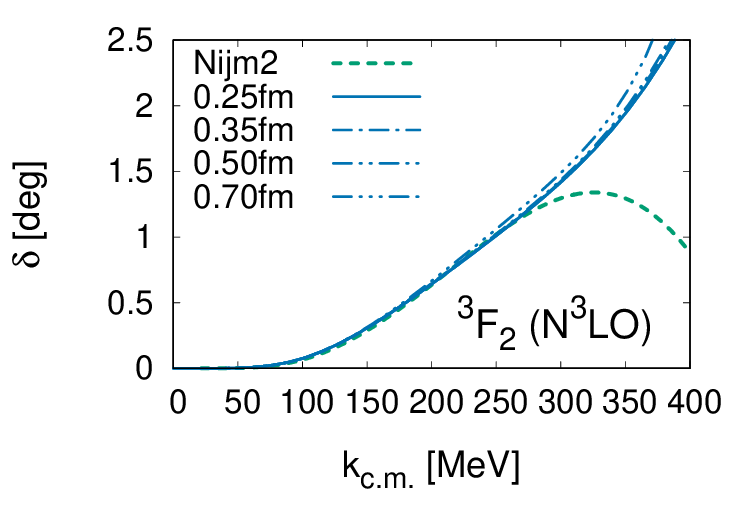}
    \includegraphics[width=5.6cm]{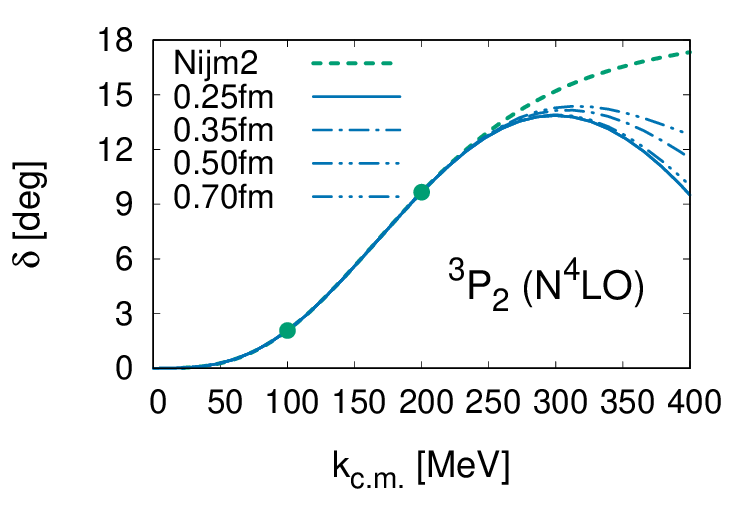}
    \includegraphics[width=5.6cm]{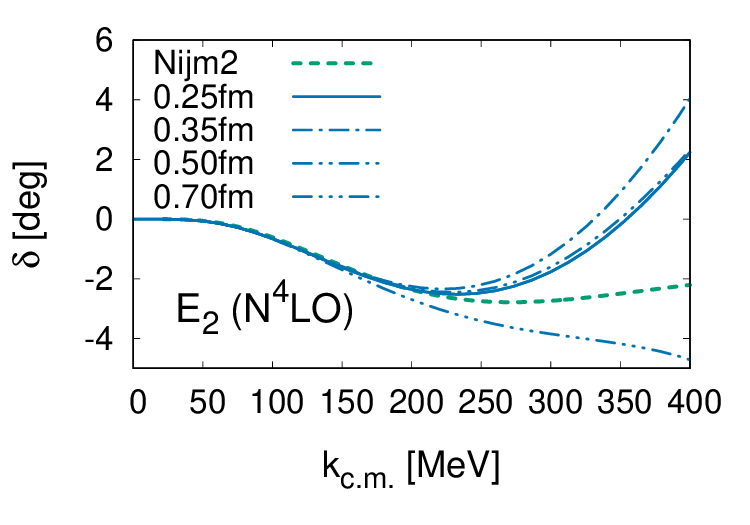}
    \includegraphics[width=5.6cm]{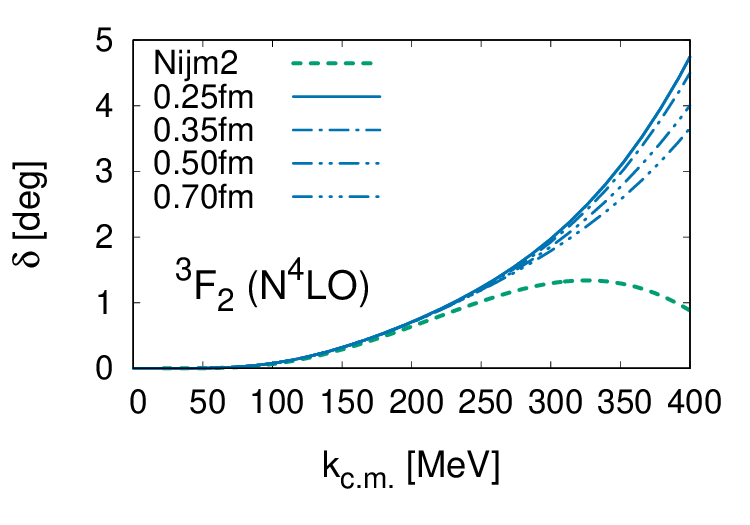}
\end{center}
  \caption{Phase shifts (nuclear bar) for the $^3S_1$-$^3D_1$
    and $^3P_2$-$^3F_2$ channels at ${\rm N^3LO}$ and ${\rm N^4LO}$
    (i.e. up to the inclusion of leading and subleading TPE,
    respectively).
    The ${\rm LO}$ contact-range interaction is regularized with the boundary
    conditions of Eqs.~(\ref{eq:uw-reg}) and (\ref{eq:orthogonality-mod})
    at the cutoff radius $R_c$,
    while the subleading orders use the delta-shell regularization and
    couplings of Eqs.~(\ref{eq:delta-shell-coupled}) and
    (\ref{eq:contacts-coupled}).
    The finite-range interaction is always regularized with a sharp cutoff
    in coordinate space (that is, the potential is zero for $r < R_c$).
    The ${\rm LO}$ renormalization condition is the reproduction of the $^3S_1$
    ($^3P_2$) scattering length (volume), $a_1({}^3S_1) = 5.419\,{\rm fm}$
    ($a_1({}^3P_2) = -0.040\,{\rm fm}^3$), while at subleading orders
    the calculation reproduces the $^3S_1$ ($^3P_2$) nuclear bar phase
    shifts of the Nijmegen II potential at
    $k_{\rm cm} = 100\,{\rm MeV}$ and $200\,{\rm MeV}$ (these two data
    points are represented as circles in the $^3S_1$ and $^3P_2$ plots).
    The $^3S_1$-$^3D_1$ ($^3P_2$-$^3F_2$) phase shifts are calculated at
    $R_c = 0.15$, $0.25$, $0.35$ and $0.50\,{\rm fm}$
    ($R_c = 0.25$, $0.35$, $0.50$ and $0.70\,{\rm fm}$)
    and are compared with the Nijmegen II ones.
}
\label{fig:3C1-3C2}
\end{figure*}

With the previous choices one obtains the phase shifts shown
in Fig.~\ref{fig:3C1-3C2} for the $^3S_1$-$^3D_1$ and
$^3P_2$-$^3F_2$ channels.
A few comments are in order:
\begin{itemize}
\item[(i)] In the convention I use here the OPE potential enters at order
  $\nu = -1$ (because it is promoted by one order with respect to NDA by
  the identification of a low energy scale within
  the OPE potential~\cite{Birse:2005um}),
  while leading and subleading TPE remain at $\nu = 2$, $3$,
  i.e. their NDA counting.
  Thus, calculations up to order $\nu$ are labeled as ${\rm N^{(\nu+1)}LO}$,
  or equivalently, calculations including leading and subleading TPE
  are ${\rm N^3LO}$ and ${\rm N^4LO}$, respectively.

\item[(ii)] In contrast,
  Refs.~\cite{Long:2011qx,Long:2011xw,Thim:2024yks} consider
  that this low energy scale is also implicitly included in TPE, resulting
  in a promotion of leading and subleading TPE by one order and hence
  calculations including them are ${\rm N^2LO}$ and ${\rm N^3LO}$,
  respectively.
  In practical terms, for the calculations included in the present manuscript
  this is merely a change in the naming convention (unless one is explicitly
  considering the EFT truncation errors, in which case their size will be
  different).
  
\item[(iii)] It is interesting to stress that for fixing the value of
  the $^3S_1$ and $^3P_2$ scattering length and volume the direction of
  integration of the Schr\"odinger equation is downwards,
  i.e. from $r \to \infty$ to $r = R_c$.
  This generates diverging exponentials in the repulsive eigenchannels
  that are later eliminated at the cutoff radius by means of
  the regularity condition, Eq.~(\ref{eq:uw-reg}).
  However, this procedure eventually becomes numerically unstable
  if the cutoffs is very small. Indeed, numerical noise begins to appear
  at $R_c < 0.14\,{\rm fm}$ and $0.18\,{\rm fm}$
  in the $^3S_1$-$^3D_1$ and $^3P_2$-$^3F_2$
  channels, respectively.
  This is the reason why I have chosen slightly different sets of cutoffs
  for each of the coupled triplets.

\item[(iv)] The obtained phase shifts are relatively good, and seem to be
  competitive with perturbative calculations with one more
  counterterm per coupled triplet~\cite{Long:2011qx,Long:2011xw,Thim:2024yks}.
  This is still the case if the comparison is made with fully
  non-perturbative calculations in NDA (i.e. Weinberg
  counting)~\cite{Epelbaum:2003xx,Gezerlis:2014zia},
  which contain three (one) counterterms in the
  $^3S_1$-$^3D_1$ ($^3P_2$-$^3F_2$) channel.
  Yet, the aim of the present manuscript is not phenomenological success,
  but to analyze the perturbative renormalizability of
  the coupled triplets.
  That the phase shifts happen to be good is of course welcomed,
  but is immaterial in what regards their finiteness in the $R_c \to 0$ limit,
  which I further analyze in the following paragraphs.
\end{itemize}

\subsection{Cutoff dependence}

Owing to the discovery of the exceptional cutoffs~\cite{Gasparyan:2022isg},
it is advisable to explicitly check the cutoff dependence of
perturbative calculations.

In principle, if expressed in the language of subtractions in configuration
space, it is apparent after a brief analysis that distorted wave
perturbation theory converges in the $R_c \to 0$ limit.
For uncoupled channels the subleading correction
to the phase shifts reads
\begin{eqnarray}
  && - \frac{\delta^{(\nu)}}{\sin^2 \delta_{\rm LO}} =
  \frac{2\mu}{k^{2l+1}}\,\mathcal{A}^2(k)\,\hat{I}_k(R_c) \, ,
\end{eqnarray}
with $\hat{I}_k$ the perturbative integral, which after two (explicit)
subtractions can be rewritten as
\begin{eqnarray}
  \hat{I}_k(R_c) &=&
  \int_0^{\infty}\,dr\,V_F^{(\nu)}(r; R_c)\,\left[ \hat{u}_k^2 - \hat{u}_0^2 - 2\,\hat{u}_0 \hat{u}_2\,k^2 \right] \nonumber \\
  && + \lambda_{0R}(R_c) + \lambda_{2R}(R_c)\,k^2 \, , \nonumber \\
\end{eqnarray}
where the radial dependence of the reduced wave function is implicit and
with $\lambda_{0R}$ and $\lambda_{2R}$ renormalized versions of
the (non-renormalized or bare) $\lambda_{0}$ and $\lambda_{2}$
subtraction constants introduced
in Eq.~(\ref{eq:Ik-two-subs}).
In this form, the $R_c \to 0$ limit of the perturbative integral is given by
\begin{eqnarray}
  \lim_{R_c \to 0} \hat{I}_k(R_c) &=&
  \hat{I}_{k, {\rm conv}} + \lambda_{0R} + \lambda_{2R}\,k^2 \, ,
\end{eqnarray}
where each term is finite.

The previous explicit subtraction scheme, though transparent regarding the
existence of the zero (or infinite) cutoff limit, is numerically
cumbersome, particularly if extended to coupled channels.
Thus it is more practical to regularize the EFT potential and impose suitable
renormalization conditions.
If one works with the bare subtraction constants $\lambda_0$ and $\lambda_2$,
it is still apparent that the cutoff can be removed.
Indeed, the relation between the bare and renormalized constants is trivial
\begin{eqnarray}
  \lambda_0 + \int_0^{\infty}\,dr\,V_F^{(\nu)}(r; R_c)\,\hat{u}_0^2 &=&
  \lambda_{0R} \, , \\
  \lambda_2 + 2\,\int_0^{\infty}\,dr\,V_F^{(\nu)}(r; R_c)\,\hat{u}_0 \hat{u}_2 &=&
    \lambda_{2R} \, , 
\end{eqnarray}
from which it can be deduced that there is essentially no difference
between the two schemes at finite cutoffs.

Yet, if one operates with a contact-range potential instead,
then there is a possible complication:
the explicit representation of the short-range physics by means of a
regularized contact-range potential, though numerically convenient,
might accidentally generate artifacts in the cutoff
dependence of the results, a point which is explained
in much more detail in Ref.~\cite{PavonValderrama:2025tmp}.
The specific artifact I will be watching for 
are the exceptional cutoffs discovered in~\cite{Gasparyan:2022isg}.

To illustrate how this complication may arise,
it is useful to begin by defining the reminder phase shift
\begin{eqnarray}
  \delta_{C}^{(\nu)}(k) = \delta_{\rm exp}(k) - \delta_{\rm LO}(k) - \delta_F^{(\nu)}(k) -
  \sum_{\mu < \nu}\,\delta^{(\mu)}(k) \, , \nonumber \\
  \label{eq:delta-C}
\end{eqnarray}
which is basically the difference of the experimental phase shift and
all the EFT contributions to it up to order $\nu$, with the exception of
the subtractions / contacts.
This is the quantity to which the counterterms are fitted.
In the expression above, $\delta_{\rm exp}$ are the {\it experimental} phase
shifts (or, more properly, the phase shifts extracted from a partial
wave analysis such as~\cite{Stoks:1993tb} or a high-quality
potential model such as Nijmegen II~\cite{Stoks:1994wp}),
$\delta_{\rm LO}$ the ${\rm LO}$ phase shift, $\delta_F^{(\nu)}$
the order $\nu$-th finite-range contribution given by
\begin{eqnarray}
  && - \frac{\delta^{(\nu)}_F}{\sin^2 \delta_{\rm LO}} =
  \frac{2\mu}{k^{2l+1}}\,\mathcal{A}^2(k)\,
  \int_0^{\infty}\,dr\,V_F^{(\nu)}(r; R_c)\,\hat{u}_k^2(r) \, ,
  \nonumber \\
\end{eqnarray}
and $\delta^{(\mu)}$ the complete (finite- and contact-range) subleading
contribution for $\mu < \nu$.
For simplicity I have ignored loop corrections (though they do not change
the arguments below).
From the previous, if there are two counterterms they can be determined from
solving a linear system
\begin{eqnarray}
  \begin{pmatrix}
    \delta_C^{(\nu)}(k_1) \\
    \delta_C^{(\nu)}(k_2) 
  \end{pmatrix} =
                {\bf M}_C (R_c)\,
                  \begin{pmatrix}
    C_0^{(\nu)}(R_c) \\
    C_2^{(\nu)}(R_c)
  \end{pmatrix} \, ,
\end{eqnarray}
where ${\bf M}_C$ is the matrix describing the linear transformation
between counterterms and their contribution to the perturbative
phase shifts.
It is interesting to notice that this matrix depends only on the matrix
elements of the contact-range interactions when sandwiched between
the ${\rm LO}$ wave functions. Its dependence with subleading
order quantities is indirect and mediated by the regularization of
the contact-range operators multiplying the $C_{2n}^{(\nu)}$
couplings and the number of counterterms, which determines
the dimensionality of the matrix.

The complication that might appear at this point is what happens
when there is a cutoff for which the linear system
is singular, that is:
\begin{eqnarray}
  \det{{\bf M}_C (R_c^*)} = 0 \, .
\end{eqnarray}
If this is the case the counterterms diverge as
\begin{eqnarray}
  C_{2n}(R_c) \propto \frac{1}{\det{{\bf M}_C(R_c)}}
  \quad \mbox{for $R_c \to R_c^*$}\, ,
\end{eqnarray}
which is not a problem per se --- the counterterms are not observable ---
unless this divergence propagates to the phase shifts.
Even though it is not trivial to determine what types of zeros of
$\det{{\bf M}_C}$ do indeed generate divergent phase shifts,
it is nonetheless much easier to identify a particular case
in which these zeros are guaranteed to be harmless:
\begin{itemize}
\item[(i)] When a column of the perturbative matrix ${\bf M}_C$ approaches zero
as $(R_c - R_c^*)^{\alpha}$, this zero can be trivially canceled
by a counterterm diverging as $(R_c - R_c^*)^{-\alpha}$.
\end{itemize}
In this case, which corresponds to the factorizable zeros of
Ref.~\cite{Gasparyan:2022isg}, the phase shifts will
not be affected.

The uncoupled attractive triplets, when regularized with a delta-shell
contact-range interaction of the type:
\begin{eqnarray}
  V_C(r; R_c) = \left[ C_0 + C_2 k^2 \right]\,\frac{\delta(r-R_c)}{4\pi R_c^2}
  \, ,
\end{eqnarray}
provide a simple illustration of the absence of exceptional cutoffs.
If one defines for convenience a reduced reminder phase shift
\begin{eqnarray}
  \hat{\delta}_{C}^{(\nu)}(k) = -\sin^2 \delta_{\rm LO}\,
  \frac{2\mu}{k^{2l+1}}\,\mathcal{A}^2(k)\,\delta_C^{(\nu)}(k) \, ,
  \label{eq:delta-C-reduced}
\end{eqnarray}
then the renormalization condition of fixing $\hat{\delta}_C^{(\nu)}$
at two different momenta takes the form:
\begin{eqnarray}
  \begin{pmatrix}
    \hat{\delta}_C^{(\nu)}(k_1) \\
    \hat{\delta}_C^{(\nu)}(k_2) 
  \end{pmatrix} &=&
  \frac{1}{4\pi R_c^2}
  \begin{pmatrix}
    \hat{u}_{k_1}^2(R_c) & k_1^2\,\hat{u}_{k_1}^2(R_c) \\
    \hat{u}_{k_2}^2(R_c) & k_2^2\,\hat{u}_{k_2}^2(R_c)
  \end{pmatrix}\,
  \begin{pmatrix}
    C_0^{(\nu)} \\
    C_2^{(\nu)}
  \end{pmatrix} \nonumber \\
  &=&  \hat{{\bf M}}_C(R_c)\,
  \begin{pmatrix}
    C_0^{(\nu)} \\
    C_2^{(\nu)}
  \end{pmatrix} \, ,
  \label{eq:perturbative-linear-system}
\end{eqnarray}
for which the determinant of this (reduced) perturbative matrix is
equal to
\begin{eqnarray}
  \det{\hat{{\bf M}}_C(R_c)} = (k_2^2 - k_1^2)
  \,\frac{\hat{u}_{k_1}^2(R_c)\,\hat{u}_{k_2}^2(R_c)}{16\pi^2\,R_c^4}
  \, .
\end{eqnarray}
Provided that the zeros of the ${\rm LO}$ reduced wave function are
independent of the momentum, i.e.
\begin{eqnarray}
  \hat{u}_{k_1}(R_c) = 0 \quad \mbox{if and only if}
  \quad \hat{u}_{k_2}(R_c) = 0 \, , \label{eq:zero-condition}
\end{eqnarray}
there will be no exceptional cutoffs.
If one is renormalizing the ${\rm LO}$ reduced wave function
with a boundary condition, then the condition
of Eq.~(\ref{eq:zero-condition}) is a consequence
of the orthogonality constraint at the cutoff radius
\begin{eqnarray}
  W(\hat{u}_{k_1}, \hat{u}_{k_2}){\Big|}_{R_c} = 0 \, ,
  \label{eq:Wronskian-uncoupled}
\end{eqnarray}
that is, the uncoupled channel version of
Eq.~(\ref{eq:Wronskian-coupled}).
It is worth noticing that a ${\rm LO}$ energy- and momentum-independent
contact-range interaction regularized with a delta-shell also
generates wave functions fulfilling the condition of momentum
independence of the zeros of the wave function, i.e.
Eq.~(\ref{eq:zero-condition}),
as explained in~\cite{PavonValderrama:2025tmp}.
In either of these two regularizations, for $R_c \to R_c^*$ one has
\begin{eqnarray}
  \det{{\hat{\bf M}}_C(R_c)} \propto (R_c - R_c^*)^4 \, ,
  \label{eq:det-quartic-zero}
\end{eqnarray}
and
\begin{eqnarray}
  C_{2n}(R_c) \propto \frac{1}{(R_c - R_c^*)^2} \, .
\end{eqnarray}
Yet, if one rewrites the contact-range interaction in terms of equivalent
subtraction constants
\begin{eqnarray}
  \lambda_{2n}(R_c) = C_{2n}(R_c)\,\frac{\hat{u}^2(R_c)}{4 \pi R_c^2} \, ,
\end{eqnarray}
where the residual $k$-dependence of the reduced wave function has been ignored
for simplicity, then it is apparent that $\lambda_{2n}$ is a smooth
function of the cutoff at $R_c = R_c^*$.
From this the absence of exceptional cutoffs becomes evident.

\begin{figure}[ttt]
  \begin{center}
    \includegraphics[width=7.5cm]{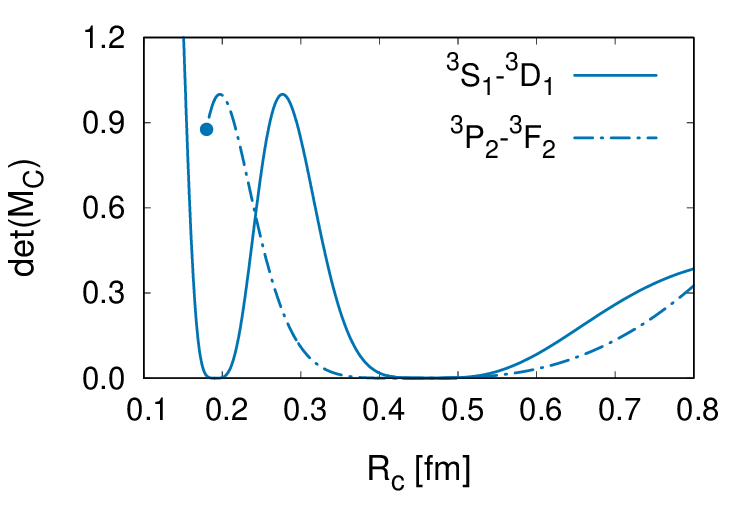}
\end{center}
  \caption{Perturbative determinant for the $^3S_1$-$^3D_1$ and $^3P_2$-$^3F_2$
    partial waves and the renormalization conditions used
    in Fig.~\ref{fig:3C1-3C2}.
    The units have been normalized as to generate local maxima of one
    in the region between the first and second zero of
    the perturbative determinant (where for the $^3P_2$-$^3F_2$
    case the second zero is not actually reached,
    but just implied).
    These zeros are quartic and correspond to the appearance of
    deeply bound states.
    The point calculated with the lowest numerically stable cutoff
    is indicated with a circle (provided it falls within
    the plot range).
}
\label{fig:det}
\end{figure}

For the case at hand, i.e. coupled channels, it is however
less straightforward to guarantee the absence of exceptional cutoffs.
The complication is that there are two linearly independent solutions,
which for instance generate a more involved relation between
subtraction constants and counterterms:
\begin{eqnarray}
  \lambda_{2n, \alpha \beta}(R_c) =
  C_{2n}(R_c)\,\frac{\hat{u}_{\alpha}(R_c) \hat{u}_{\beta}(R_c)}{4 \pi R_c^2} \, ,
\end{eqnarray}
which is specific for the chosen form of the contact-range potential
in Eq.~(\ref{eq:contacts-coupled}).
Yet, provided the following condition is met
\begin{eqnarray}
  \hat{u}_{k_1, \alpha}(R_c) = 0 \quad \mbox{if and only if} \quad
  \hat{u}_{k_2, \beta}(R_c) = 0 \, , \nonumber \\
\end{eqnarray}
there will be no problem.
It happens that this condition is a direct consequence of the combined
regularity and orthogonality condition of Eq.~(\ref{eq:orthogonality-mod}),
thus precluding the existence of exceptional cutoffs.
Yet, for peace of mind, it makes no harm to numerically calculate
the perturbative determinant for the specific renormalization
conditions I am using (i.e. reproducing the $^3S_1$ or $^3P_2$ nuclear
bar phase shifts at center-of-mass momenta $k = 100$ and $200\,{\rm MeV}$),
which is done in Fig.~\ref{fig:det}.
There it can be appreciated that the zeros of the determinant are quartic,
just as expected from Eq.~(\ref{eq:det-quartic-zero}).

\begin{figure*}[ttt]
  \begin{center}
    \includegraphics[width=5.6cm]{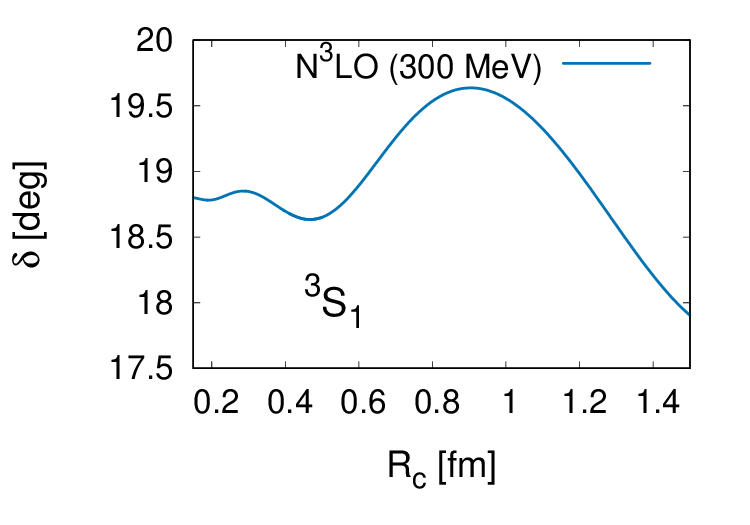}
    \includegraphics[width=5.6cm]{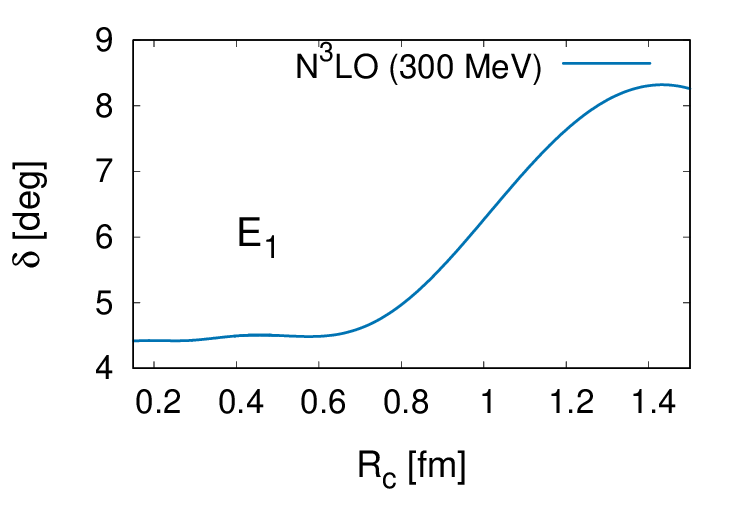}
    \includegraphics[width=5.6cm]{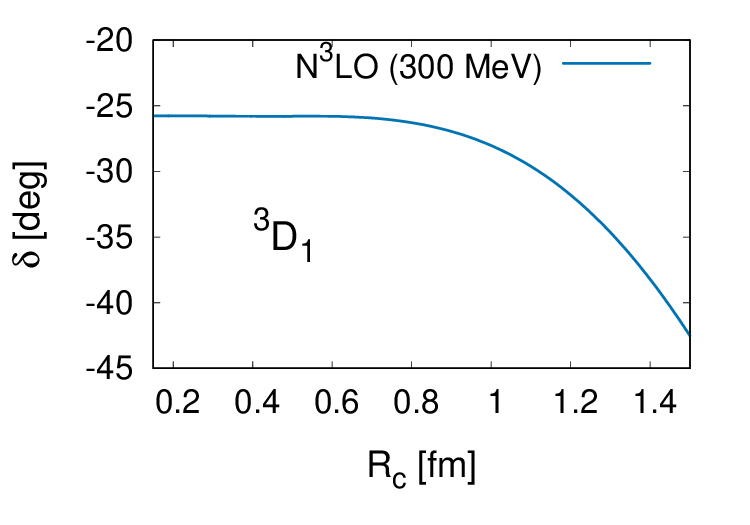}
    \includegraphics[width=5.6cm]{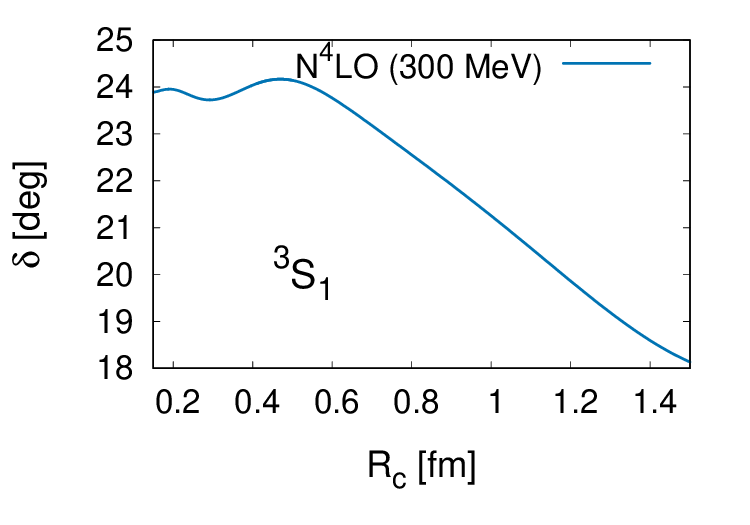}
    \includegraphics[width=5.6cm]{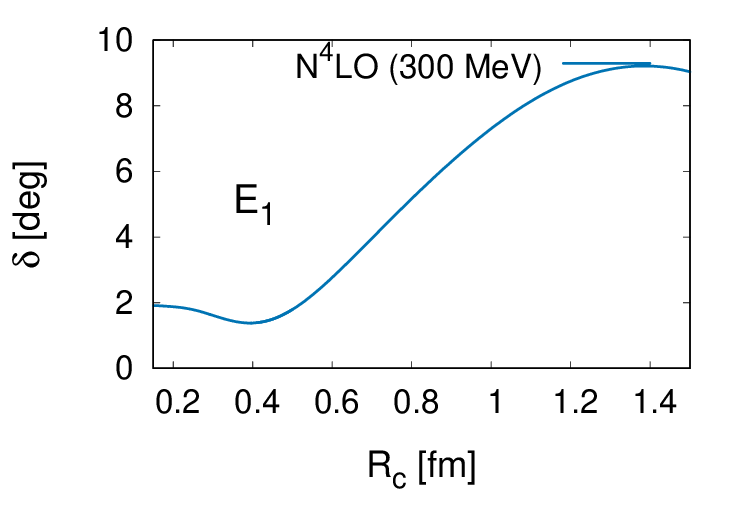}
    \includegraphics[width=5.6cm]{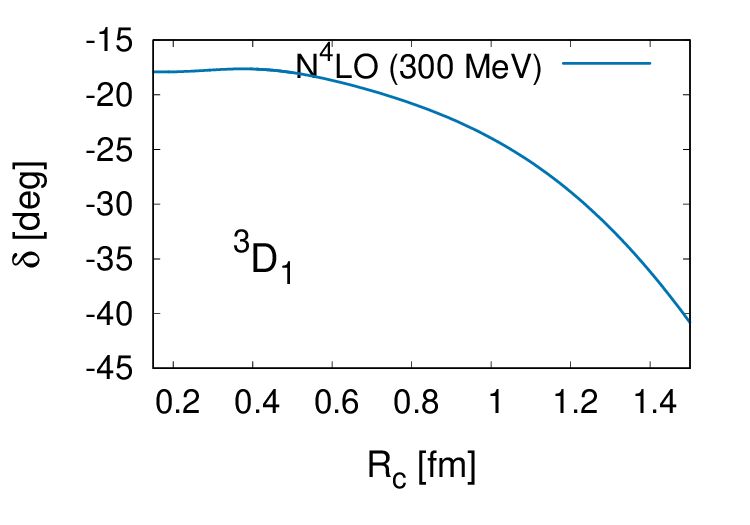}
    \includegraphics[width=5.6cm]{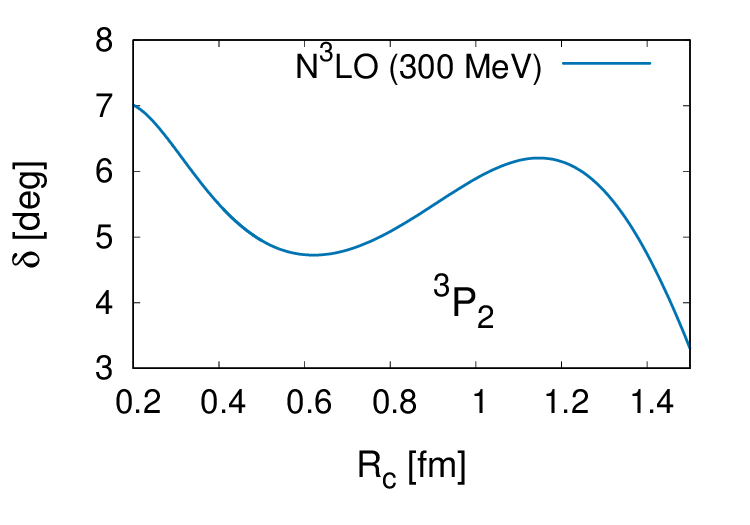}
    \includegraphics[width=5.6cm]{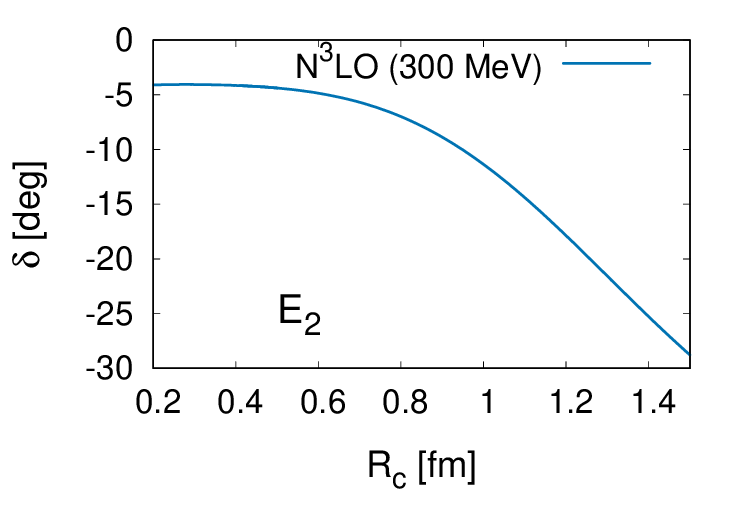}
    \includegraphics[width=5.6cm]{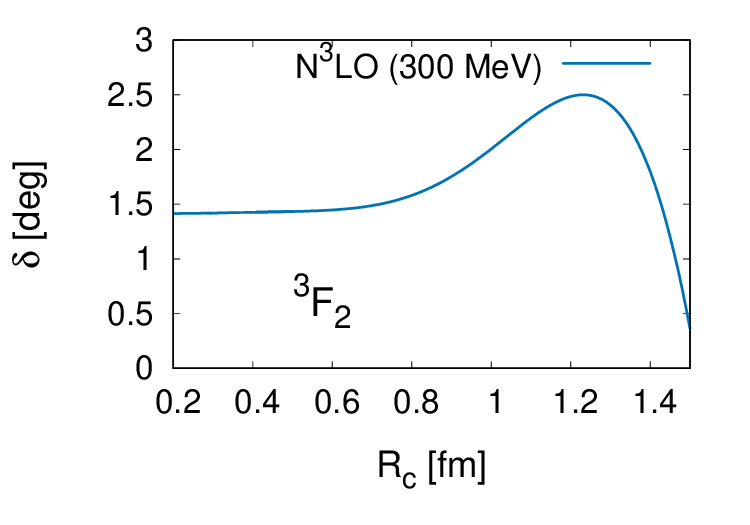}
    \includegraphics[width=5.6cm]{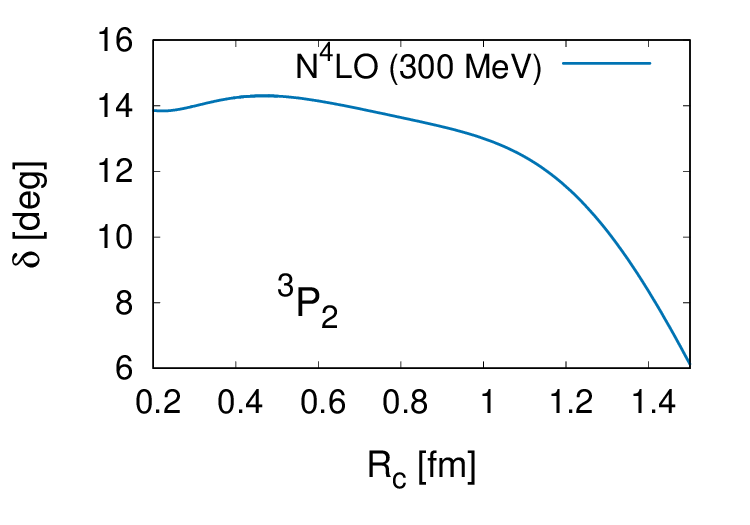}
    \includegraphics[width=5.6cm]{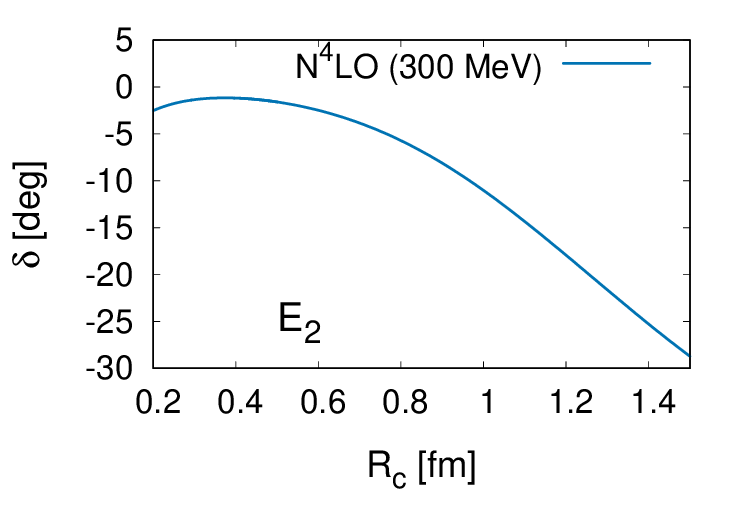}
    \includegraphics[width=5.6cm]{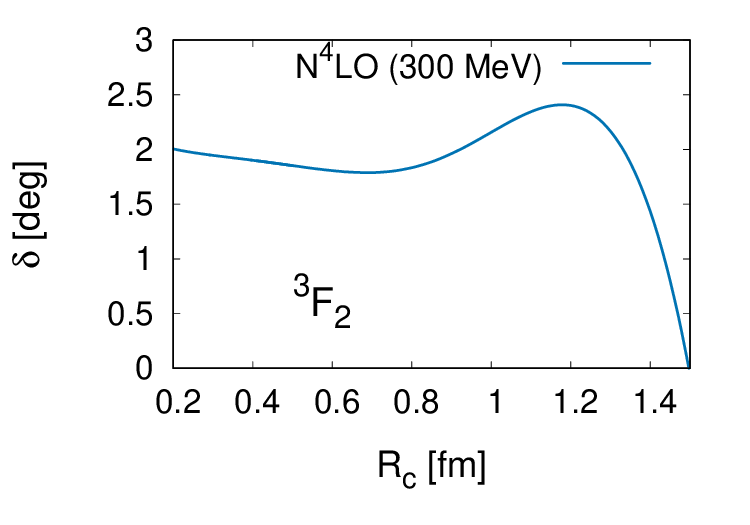}
  \end{center}
    \caption{Cutoff dependence of the $^3S_1$-$^3D_1$ and $^3P_2$-$^3F_2$
    (nuclear bar) phase shifts at  ${\rm N^3LO}$ and ${\rm N^4LO}$,
    calculated for a center-of-mass
    momentum of $k_{\rm cm} = 300\,{\rm MeV}$.
    The regularization and renormalization choices are identical to the ones
    in Fig.~\ref{fig:3C1-3C2}.
    It can be appreciated that there is convergence with respect to the cutoff
    when $R_c \to 0$, though owing to the appearance of numerical instabilities
    at hard cutoffs (which are a consequence of the existence of a repulsive
    eigenchannel for the OPE potential) the calculations
    are limited to $R_c \geq 0.14\,{\rm fm}$ and $R_c \geq 0.18\,{\rm fm}$
    in the $^3S_1$-$^3D_1$ and $^3P_2$-$^3F_2$ channels, respectively.
}
\label{fig:3C1-3C2-Rc}
\end{figure*}

Finally, the explicit cutoff dependence for the $^3S_1$-$^3D_1$ and
$^3P_2$-$^3F_2$ phase shifts at $k = 300\,{\rm MeV}$ is shown
in Fig.~\ref{fig:3C1-3C2-Rc}, where it is apparent that convergence
with respect to the cutoff is smooth and that
there are no exceptional cutoffs (though the proof of this feature
rest in the behavior of the cutoff dependence of
the perturbative determinant).

\subsection{Derivative contact-range interactions}

The analysis of the perturbative integrals and their renormalization by means of
subtractions proves the existence of the infinite (or zero) cutoff limit of
distorted wave perturbation theory.
This has the interesting implication that exceptional cutoffs are not
a problem of the renormalization process itself but of
the chosen regularization procedure.
Their appearance depends on the specific representation of the short-range
physics, i.e. on how one regularizes the contact-range potential.

Yet, regularized contact-range potentials remain the most practical
method by which to do calculations within nuclear EFT.
In this regard,
the choice of an energy-dependent delta-shell for regularizing
the contact-range potential is not driven by necessity
but by convenience.
It simply provides the easiest way to regularize and renormalize
the coupled triplets for the specific boundary
conditions of Eqs.~(\ref{eq:uw-reg}) and
(\ref{eq:orthogonality-mod}).

Though energy- and momentum-dependence do in principle generate equivalent
EFT observables (modulo higher order uncertainties), the practical
application of EFT often prefers momentum-dependent contacts,
few-body calculations being a good example~\footnote{Yet, it is worth noticing
  that the main obstacle for extending the current analysis to the few-body case
  is not whether one uses energy- or momentum-dependent contacts but
  developing a detailed understanding of the short-range behavior of
  the few-body wave functions}.
Thus, it is interesting to check whether one can construct contacts with
derivatives that do not display exceptional cutoffs.
Besides, as noted in~\cite{Gasparyan:2022isg}, for energy-dependent contacts
the zeros of the perturbative determinant are often (though not always)
factorizable, in which case calculations do not show exceptional cutoffs.

Curiously, for boundary condition regularization, energy- and momentum-dependent
counterterms are exactly equivalent~\cite{PavonValderrama:2025tmp}.
For uncoupled attractive triplets this can be easily understood by considering
the explicit derivative contact-range interaction:
\begin{eqnarray}
  V_C(r; R_c) = \left[ C_0 + C_2\,\,\hat{D}_2 \right]\,
  \frac{\delta(r-R_c)}{4\pi R_c^2}
  \, , \label{eq:D2-p-space}
\end{eqnarray}
with $\hat{D}_2$ an operator containing up to two-derivatives,
where its structure follows the general pattern
\begin{eqnarray}
  \hat{D}_2 =
  -\frac{d^2}{dr^2} + \frac{a_1}{r}\,\frac{d}{dr} + \frac{a_0}{r^2} \, ,
\end{eqnarray}
with $a_0$ and $a_1$ numerical constants that depend
on the particular choice of derivative operator.
The matrix elements of this derivative interaction are
\begin{eqnarray}
  \langle V_C \rangle &=&
  \frac{C_0(R_c)}{4\pi R_c^2}\,u^2(R_c)
  + \frac{C_2(R_c)}{4\pi R_c^2}\,D_2(u^2(r)) \Big|_{r = R_c} \, ,
  \nonumber \\
\end{eqnarray}
with $D_2$ the derivative operator
\begin{eqnarray}
  D_2 = -\frac{d^2}{dr^2} - \frac{a_1}{r}\,\frac{d}{dr} +
  \frac{a_0 + a_1}{r^2} \, , \label{eq:D2-r-space}
\end{eqnarray}
which is obtained from integrating by parts the $\hat{D}_2$ operator
originally acting on the delta-shell.

By writing down the linear system by which one calibrates the values of
the $C_0^{(\nu)}$ and $C_2^{(\nu)}$ couplings, one arrives at
\begin{eqnarray}
  \begin{pmatrix}
    \hat{\delta}_C^{(\nu)}(k_1) \\
    \hat{\delta}_C^{(\nu)}(k_2) 
  \end{pmatrix} &=&
  \frac{1}{4\pi R_c^2}
  \begin{pmatrix}
    \hat{u}_{k_1}^2(R_c) & (D_2\hat{u}_{k_1}^2)(R_c) \\
    \hat{u}_{k_2}^2(R_c) & (D_2\,\hat{u}_{k_2}^2)(R_c)
  \end{pmatrix}\,
  \begin{pmatrix}
    C_0^{(\nu)} \\
    C_2^{(\nu)}
  \end{pmatrix} \nonumber \\
  &=&  \hat{{\bf M}}_C^{\rm d}(R_c)\,
  \begin{pmatrix}
    C_0^{(\nu)} \\
    C_2^{(\nu)}
  \end{pmatrix} \, ,
\end{eqnarray}
with $\hat{\delta}_C$ as defined in Eq.~(\ref{eq:delta-C-reduced}) and
where $D_2 \hat{u}_k^2$ can be written as
\begin{eqnarray}
  (D_2 \hat{u}_k^2)(R_c) = \left[ h_k(R_c) + 2\,k^2 \right]\,\hat{u}_k^2(R_c) \, ,
\end{eqnarray}
with the function $h_k$ given by
\begin{eqnarray}
  h_k(R_c) &=& \frac{(a_0+a_1)}{R_c^2} - \frac{2 a_1}{R_c}\,d_k(R_c) \nonumber \\
  &-& 2 \,d_k^2 (R_c) - 2\,\left[ 2\mu\,V_{\rm LO}(R_c) + \frac{l(l+1)}{R_c^2}
    \right]
     \, ,
  \nonumber \\
\end{eqnarray}
where $d_k$ is the logarithmic derivative at $r = R_c$
\begin{eqnarray}
  d_k(R_c) = \frac{\hat{u}_k'(R_c)}{\hat{u}_k(R_c)} \, .
\end{eqnarray}

For boundary condition regularization the orthogonality condition 
is equivalent to the Wronskian condition of Eq.~(\ref{eq:Wronskian-uncoupled}),
which in turn can be rewritten as
\begin{eqnarray}
  \frac{\hat{u}_{k_1}'(R_c)}{\hat{u}_{k_1}(R_c)} =
  \frac{\hat{u}_{k_2}'(R_c)}{\hat{u}_{k_2}(R_c)} \, .
\end{eqnarray}
This implies that $d_k$ is energy independent, $d_k(R_c) = d(R_c)$.
Therefore $h_k$ is also energy independent
\begin{eqnarray}
  h_k(R_c) = h(R_c) \, .
\end{eqnarray}
This leads to the following simplification for the linear system
\begin{eqnarray}
&&
  \begin{pmatrix}
    \hat{\delta}_C^{(\nu)}(k_1) \\
    \hat{\delta}_C^{(\nu)}(k_2) 
  \end{pmatrix} =
  \hat{{\bf M}}_C^{\rm d}(R_c)\,
  \begin{pmatrix}
    C_0^{(\nu)} \\
    C_2^{(\nu)}
  \end{pmatrix} \nonumber \\
  && \,\, = 
    \frac{1}{4\pi R_c^2}
  \begin{pmatrix}
    \hat{u}_{k_1}^2(R_c) & k_1^2\,\hat{u}_{k_1}^2(R_c) \\
    \hat{u}_{k_2}^2(R_c) & k_2^2\,\hat{u}_{k_2}^2(R_c)
  \end{pmatrix}\,
  \begin{pmatrix}
    C_0^{(\nu)} + h\,C_2^{(\nu)} \\
    2\,C_2^{(\nu)}
  \end{pmatrix} \nonumber \\
  && \,\, = \hat{{\bf M}}_C(R_c)\,
    \begin{pmatrix}
    C_0^{(\nu)} + h\,C_2^{(\nu)} \\
    2\,C_2^{(\nu)}
  \end{pmatrix}
\end{eqnarray}
which is simply the linear system of Eq.~(\ref{eq:perturbative-linear-system})
for the energy-dependent contact, except for a linear redefinition
of the couplings (which are not observable).

Thus, it is apparent that for boundary condition regularization energy-
and momentum-dependence are exactly equivalent for every cutoff.
For other regulator choices, provided exceptional cutoffs are absent,
this exact equivalence is limited to the $R_c \to 0$ limit.
This is shown in~\cite{PavonValderrama:2025tmp} when the ${\rm LO}$ wave
function is renormalized with a delta-shell instead of
with boundary conditions, in which case the cutoff dependence of
the phase shifts happens to be different.
Convergence in the $R_c \to 0$ limit is still unique, as expected for
results that have been properly renormalized.

The argument for the coupled channel case is analogous, except for
the complications derived from the existence of attractive and
repulsive eigenchannels at short distances.
For including derivatives one simply writes the contact-range potential as
\begin{eqnarray}
  {\bf V}_{C}(r; R_c) = \left[ {\bf C}_0(R_c) +
    {\bf C}_2(R_c)\,\hat{D}_2 \right]\,
  \frac{\delta(r-R_c)}{4 \pi R_c^2} \, ,  \nonumber \\
  \label{eq:delta-shell-coupled-deriv}
\end{eqnarray}
with $\hat{D}_2$ a derivative operator of the type defined
in Eq.~(\ref{eq:D2-p-space}).
Once derivatives are involved the angular momentum structure of
the ${\bf C}_{2n}$ couplings is more constrained than before.
If one wants to prevent the appearance of exceptional cutoffs,
it is advisable to follow the pattern
\begin{eqnarray}
  {\bf C}_{2n} =
  C_{2n}
  \begin{pmatrix}
    1 & \frac{1}{f_j} \\
    \frac{1}{f_j} & \frac{1}{f_j^2}
  \end{pmatrix} \, ,
\end{eqnarray}
which is tied to the boundary condition of Eq.~(\ref{eq:uw-reg}),
where $f_j$ was also defined.
The matrix elements of this type of counterterm only involve the part of
the wave function that corresponds to the attractive eigenchannel at
short enough distances, which is given by:
\begin{eqnarray}
  v_{A k\rho}(r) = u_{k\rho}(r) + \frac{1}{f_j}\,w_{k\rho}(r) \, ,
\end{eqnarray}
with $\rho \in \{ \alpha, \beta \}$.
With this definition, one obtains
\begin{eqnarray}
  \langle V_C \rangle_{\rho \sigma} &=&
  \frac{C_0(R_c)}{4\pi R_c^2}\,v_{A k \rho}(R_c)\,v_{A k \sigma}(R_c) \nonumber \\
  &+& \frac{C_2(R_c)}{4\pi R_c^2}\,
  D_2(v_{A k \rho}\,v_{A k \sigma}) \Big|_{r = R_c} \, ,
\end{eqnarray}
with $D_2$ the derivative operator defined in Eq.~(\ref{eq:D2-r-space})
and $\rho, \sigma \in \{ \alpha, \beta \}$.
If one now changes the normalization of the wave functions to the ``hat''
normalization of Eq.~(\ref{eq:hat-normalization}),
calculates the matrix elements
\begin{eqnarray}
  \langle 1 \rangle_{\rho \sigma}(k) &=& \hat{v}_{A k\rho}\,\hat{v}_{A k\sigma} \Big|_{r=R_c}
  \, , \\
  \langle \hat{D}_2 \rangle_{\rho \sigma}(k) &=& D_2(\hat{v}_{A k\rho}\,\hat{v}_{A l\sigma})\,
  \Big|_{r=R_c} \, ,
\end{eqnarray}
and defines the perturbative matrix as
\begin{eqnarray}
  \hat{{\bf M}}_{C\rho\sigma}^{d}(R_c) =
  \begin{pmatrix}
    \langle 1 \rangle_{\rho \sigma}(k_1) &
    \langle \hat{D}_2 \rangle_{\rho \sigma}(k_1) \\
    \langle 1 \rangle_{\rho \sigma}(k_2) &
    \langle \hat{D}_2 \rangle_{\rho \sigma}(k_2)
  \end{pmatrix} \, ,
\end{eqnarray}
then, it is possible to determine the two couplings by solving any of
the following linear systems (or two arbitrary linearly independent
combinations of the equations below):
\begin{eqnarray}
    \begin{pmatrix}
    \hat{\delta}_{C \alpha}^{(\nu)}(k_1) \\
    \hat{\delta}_{C \alpha}^{(\nu)}(k_2) 
  \end{pmatrix}
  &=&  \hat{{\bf M}}_{C\alpha \alpha}^d(R_c)\,
  \begin{pmatrix}
    C_0^{(\nu)} \\
    C_2^{(\nu)}
  \end{pmatrix} \, , \\
      \begin{pmatrix}
    \hat{\delta}_{C \beta}^{(\nu)}(k_1) \\
    \hat{\delta}_{C \beta}^{(\nu)}(k_2) 
  \end{pmatrix}
  &=&  \hat{{\bf M}}_{C\beta \beta}^d(R_c)\,
  \begin{pmatrix}
    C_0^{(\nu)} \\
    C_2^{(\nu)}
  \end{pmatrix} \, , \\
      \begin{pmatrix}
    \hat{\epsilon}_{C}^{(\nu)}(k_1) \\
    \hat{\epsilon}_{C}^{(\nu)}(k_2) 
  \end{pmatrix}
  &=&  \hat{{\bf M}}_{C\alpha \beta}^d(R_c)\,
  \begin{pmatrix}
    C_0^{(\nu)} \\
    C_2^{(\nu)}
  \end{pmatrix} \, , 
\end{eqnarray}
with
\begin{eqnarray}
  \hat{\delta}_{C \alpha}^{(\nu)}(k) &=& -\sin^2 \delta_{\alpha,\rm LO}\,
  \frac{2\mu}{k^{2j-1}}\,\mathcal{A}_{\alpha}^2(k)\,\delta_{C \alpha}^{(\nu)}(k) \, ,
  \\
  \hat{\delta}_{C \beta}^{(\nu)}(k) &=& -\sin^2 \delta_{\beta,\rm LO}\,
  \frac{2\mu}{k^{2j+3}}\,\mathcal{A}_{\beta}^2(k)\,\delta_{C \beta}^{(\nu)}(k) \, ,
  \\
  \hat{\epsilon}_{C}^{(\nu)}(k) &=& \left( \frac{1}{\cot{\delta_{\alpha, \rm LO}} -
    \cot{\delta_{\beta, \rm LO}}} \right)\, \nonumber \\
  && \,\times\,
  \frac{2\mu}{k^{2j+1}}\,\mathcal{A}_{\alpha}(k)\,\mathcal{A}_{\beta}(k)\,
  \epsilon_{C}^{(\nu)}(k) \, ,
\end{eqnarray}
where $\delta_{C \alpha}^{(\nu)}$, $\delta_{C \beta}^{(\nu)}$ and $\epsilon_{C}^{(\nu)}$
are defined as in Eq.~(\ref{eq:delta-C}).
Even though the linear systems above have been defined with respect to
the perturbative corrections to the eigen phase shifts,
this is generalizable to the nuclear bar phase shifts.
Indeed, the perturbative corrections to the nuclear bar phase shifts (or any
other parametrization of the phase shifts for that matter) are linear
combinations of the ones for the eigen phase shifts.
I have only chosen the eigen representation because it allows for a more
transparent analysis.

The interesting thing to notice is that for boundary condition regularization
one ends up with
\begin{eqnarray}
  \frac{\langle \hat{D}_2 \rangle_{\rho \sigma}(k)}
       {\langle 1 \rangle_{\rho \sigma}(k)}
  &=& h(R_c) + 2\,k^2 \, ,
  \label{eq:D2-matrix-elements}
\end{eqnarray}
with $h$ not only energy-independent, but also independent on the particular
choice of asymptotic solution.
The complete expansion of $h$ is in principle
\begin{eqnarray}
  h_{k ,\rho \sigma}(R_c) &=& \frac{(a_0+a_1)}{R_c^2} - \frac{a_1}{R_c}\,
  \left[ d_{k,A \rho}(R_c) + d_{k, A \sigma}(R_c) \right] \nonumber \\
  &-& 2 \,d_{k, A \rho} (R_c)\,d_{k, A \sigma} (R_c) -
  4\mu\,V_{\rm LO, AA}(R_c) \nonumber \\
  &-&
  2\,\frac{(j-1)\,j}{R_c^2}\,\frac{u_{k\rho} u_{k\sigma}}{v_{A k\rho} v_{A k\sigma}}\,\Big|_{r = R_c}
  \nonumber \\
  &-&
  2\,\frac{(j+1)(j+2)}{R_c^2}\,\frac{w_{k\rho} w_{k\sigma}}{v_{A k\rho} v_{A k\sigma}}\,\Big|_{r = R_c}
     \, ,
\end{eqnarray}
where $V_{\rm LO, AA}$ refers to the ${\rm LO}$ potential in the attractive
eigenchannel.
The expression above depends on $k$, $\rho$ and $\sigma$,
with $d_{k, A \rho}$ given by
\begin{eqnarray}
  d_{k, A \rho}(R_c) = \frac{v_{A k \rho}'}{v_{A k \rho}} \Big|_{r = R_c} \, .
\end{eqnarray}
Yet, for the specific boundary condition of Eq.~(\ref{eq:uw-reg})
one has
\begin{eqnarray}
  d_{k, A \rho}(R_c) &=&
  \frac{u_{k\rho}' +\frac{1}{f_j} w_{k\rho}'}
       {u_{k\rho} +\frac{1}{f_j} w_{k\rho}} \Big|_{r = R_c} \nonumber \\
       &=& \frac{f_j}{1 + f_j^2}\,
       \frac{w_{k\rho}' + f_j\,u_{k\rho}'}{u_{k\rho}} \Big|_{r = R_c} \, ,
\end{eqnarray}
where in the second line the orthogonality
condition of Eq.~(\ref{eq:orthogonality-mod}) appears,
which implies in turn that there is no dependence on $k$ or $\rho$,
i.e. $d_{k, A \rho} = d_A$.
Conversely, for the terms involving the orbital angular momentum one has
\begin{eqnarray}
  \frac{u_{k\rho}(R_c)}{v_{A k\rho}(R_c)} &=& \frac{1}{1 + f_j^2} \, , \\
  \frac{w_{k\rho}(R_c)}{v_{A k\rho}(R_c)} &=& \frac{f_j}{1 + f_j^2} \, , 
\end{eqnarray}
which is again independent of $k$ or $\rho$.
As a consequence, for the boundary condition of Eq.~(\ref{eq:uw-reg}),
one indeed has $h_{k, \rho \sigma}(R_c) = h(R_c)$,
proving Eq.~(\ref{eq:D2-matrix-elements}).

Thus, with the boundary condition regularization of Eq.~(\ref{eq:uw-reg}),
each of the linear systems corresponding to each type of phase shift
reduces to the common form
\begin{eqnarray}
  \hat{{\bf M}}_{C\rho\sigma}^{d}(R_c) =
  \langle 1 \rangle_{\rho \sigma} \,
  \begin{pmatrix}
    1 & h + 2\,k_1^2 \\
    1 & h + 2\,k_2^2 
  \end{pmatrix} \, ,
\end{eqnarray}
where I have explicitly taken into account that
$\langle 1 \rangle_{\rho \sigma}(k)$ becomes energy independent
for this regularization.
This form basically means that:
\begin{itemize}
\item[(i)] Given two arbitrary renormalization conditions, there will always
  be a solution (and thus no exceptional cutoffs).
\item[(ii)] Owing to the linear transformation
  \begin{eqnarray}
    && \langle 1 \rangle_{\rho \sigma}\,
    \begin{pmatrix}
      1 & h + 2\,k_1^2 \\
      1 & h + 2\,k_2^2 
    \end{pmatrix}\,
    \begin{pmatrix}
      C_0^{(\nu)} \\
      C_2^{(\nu)} 
    \end{pmatrix} \nonumber \\
    && \,\, =
    \langle 1 \rangle_{\rho \sigma}\,
    \begin{pmatrix}
      1 & k_1^2 \\
      1 & k_2^2 
    \end{pmatrix}\,
    \begin{pmatrix}
      C_0^{(\nu)} + h\,C_2^{(\nu)}\\
      2\,C_2^{(\nu)} 
    \end{pmatrix} \, ,
  \end{eqnarray}
  the derivative contact-range potential of
  Eq.~(\ref{eq:delta-shell-coupled-deriv}) yields identical results to that
  of its energy-dependent counterpart, Eq.~(\ref{eq:delta-shell-coupled}),
  where the only difference is the specific values of the couplings.
\end{itemize}
That is, the cutoff dependence of a derivative contact is identical to
that of Fig.~\ref{fig:3C1-3C2-Rc} for an energy-dependent contact (this
has been explicitly checked numerically), and as a consequence
there are no exceptional cutoffs either.

At the practical level, if one is interested in the calculation of
the phase shifts in the two-nucleon sector,
the energy-dependent representation is computationally more convenient
than the momentum-dependent one.
The reason is numerical noise, which for the contact-range potential of
Eq.~(\ref{eq:delta-shell-coupled-deriv}) is particularly
noticeable close to the cutoffs for which
the wave function has a zero.
In contrast, if one works with the energy-dependent contact of
Eq.~(\ref{eq:delta-shell-coupled}) numerical noise is almost absent,
unless one integrates the coupled-channel Sch\"odinger equation
with very low numerical accuracy.

Unfortunately, the analysis of the exceptional cutoffs presented
here for energy- and momentum-dependent contact-range
interactions is regulator specific.
While for the particular case of boundary conditions and delta-shell
regularization the absence of genuine exceptional cutoffs
can be proven analytically, this is not true in general.
Most often the only tool available is a numerical calculation of
the perturbative phase shifts in the proximity of the cutoffs
for which the perturbative matrix has a zero.
A few examples have been presented in~\cite{PavonValderrama:2025tmp}
for the uncoupled channel case.

It is not the intention of the current manuscript to extend the results
of~\cite{PavonValderrama:2025tmp} to coupled channels,
but merely to show the minimum number of counterterms
required to renormalize the coupled triplets.
Yet, the analysis of Ref.~\cite{PavonValderrama:2025tmp} indicated
that regulators that are non-local in configuration space
have genuine exceptional cutoffs.
In contrast, non-locality in momentum space is inconsequential
for the appearance of exceptional cutoffs.
However, non-local and separable regulators in momentum space generate
a certain degree of (coordinate space) non-locality at short distances,
which happens to be an unwanted consequence of their separability.
This, in turn, makes it impossible for the subleading corrections to
the phase shifts to avoid the exceptional cutoffs.
Provided one does not calculate the phase shifts close to
the exceptional cutoffs, there should be no problem though.
Alternatively, Refs.~\cite{Peng:2024aiz,Peng:2025ykg}
propose a solution to the appearance of exceptional cutoffs that exploits
the intrinsic uncertainty of EFT calculations (a solution which has been
recently tested in the three-nucleon sector~\cite{Thim:2025vhe}),
while Ref.~\cite{Yang:2024yqv} cleverly constructs a 
regulator in which exceptional cutoffs are absent.

To conclude the discussion, it is worth mentioning that the inclusion of
additional counterterms poses no problem.
The perturbative renormalizability of chiral two-pion exchange is guaranteed
after two subtractions in the perturbative integrals.
Additional subtractions do not change this fact.
If one regularizes using counterterms instead, the same issues discussed here
will appear again, with the only difference being the dimension of
the perturbative matrix.
For the specific case of boundary conditions and delta-shell regulators,
the construction of derivative contact-range potentials that
do not generate exceptional cutoffs is straightforward,
though somewhat involved, and will be left
for a future investigation.

\section{Discussion and conclusions}
\label{sec:conclusions}

Here I have reanalyzed the perturbative renormalizability of
the TPE potential in the coupled triplets in nuclear EFT
when OPE is iterated to all orders.
The question I wanted to answer is how many contacts are required to render
the perturbative corrections finite in the limit
in which the cutoff is removed ($R_c \to 0$,
as I am working in coordinate space).
I find that two counterterms are enough, which is a remarkable reduction
in their number over the previous six used to renormalize the amplitudes
in~\cite{Valderrama:2009ei,Valderrama:2011mv} or the three
proposed by Long and Yang~\cite{Long:2011xw}.
In this later case it turned out that the calculations were actually
cutoff dependent~\footnote{
  This is merely an accident of the separable momentum-space regulator
  used in~\cite{Long:2011qx,Long:2011xw}, a point which is explained
  in detail in~\cite{PavonValderrama:2025tmp}.
  Other regulators might result instead in a smooth cutoff dependence (and
  indeed they are not difficult to build in boundary condition
  regularization).
  That is, the exceptional cutoffs
  in~\cite{Long:2011xw} have nothing to do with the number of counterterms
  used in the coupled triplets (provided this number
  is larger than two).}
owing to the elusive exceptional cutoffs discovered
by Gasparyan and Epelbaum~\cite{Gasparyan:2022isg}.

The interpretation of this result --- namely that leading and
subleading TPE can be perturbatively renormalized with two counterterms ---
and its implications depends on one's own views on the relation between
renormalizability and power counting.
In what follows renormalizability will be understood as the existence of
a well-defined limit when the cutoff is removed.
Caution is advised though as this limit might not necessarily be unique
once one takes into account the existence of EFT truncation errors.
But for the sake of simplicity I will temporarily ignore this problem
and its ramifications.
Besides the previous, I will also assume the following
two propositions to be correct:
\begin{itemize}
\item[(i)] Calculations in a consistent power counting allow for
  the renormalizability of the observables.
\item[(ii)] Renormalizability alone is not enough as to fully
  determine the power counting.
\end{itemize}
In short, power counting implies renormalizability but renormalizability
does not imply power counting.
Statement (i) is used for instance by Nogga, Timmermans and
van Kolck~\cite{Nogga:2005hy} to discard NDA
in the $^3P_0$ partial wave at ${\rm LO}$, while (ii) is exemplified
by the power counting of the $^3P_1$ repulsive
triplets~\cite{Valderrama:2011mv,Long:2011xw},
which is renormalizable (in the aforementioned sense of removing
the cutoff) without the inclusion of any counterterm.
Yet, a more pedestrian example of statement (ii) is
pionless EFT~\cite{vanKolck:1998bw,Chen:1999tn},
where the iteration of contacts in the $^1S_0$ and $^3S_1$ partial waves
is not mandated by finiteness alone (after all, a theory in which
there are only perturbative interactions is perfectly renormalizable),
but by the fact that the $^1S_0$ and $^3S_1$ scattering lengths
are unnaturally large.

The previous two statements are not necessarily uncontroversial though.
Of course this is not surprising in view of the never-ending debates and
confronted opinions about renormalizability and power counting~\cite{Epelbaum:2006pt,Epelbaum:2009sd,Epelbaum:2018zli,Epelbaum:2020maf,Valderrama:2019yiv,Griesshammer:2021zzz,Gasparyan:2023rtj}.
For instance, regarding proposition (i), NDA is still defended as a valid
power counting in nuclear EFT~\cite{Epelbaum:2006pt,Epelbaum:2009sd,Epelbaum:2018zli,Epelbaum:2020maf},
which requires a different interpretation of renormalizability
as the one I have followed here.
Yet, even if one rejects the necessity of cutoff independence for hard cutoffs,
it has been recently argued that there is room for allowing modifications
to NDA~\cite{Gasparyan:2023rtj}.
Thus, I stress here that these are just the assumptions I will follow.

The bottom-line is that two counterterms is the bare minimum: even though
the phase shifts obtained with two counterterms are acceptable,
an optimal power counting probably requires more counterterms,
particularly in the $^3S_1$-$^3D_1$ channel.
In this regard it might be useful to review what previous works have done.

In the $^3P_1$ partial wave, which serves as a relevant comparison,
it is worth mentioning that in~\cite{Long:2011xw} Long and Yang
solved this issue by invoking the notion of {\it primordial
  counterterms}, i.e. counterterms originating from
the regularization of the EFT potential itself.
Even if these counterterms are unnecessary when renormalizing
the (previously regularized) EFT potential in the Schr\"odinger or
Lippmann-Schwinger equation, they do appear
when calculating the EFT potential itself.
Their point is that nuclear EFT regularization and renormalization do not
only involve the two-nucleon loops appearing when calculating the matrix
elements of the subleading EFT potential, but also the loops involving
pions in the irreducible diagrams comprising said potential.
If this is the case it could be argued that renormalizability
in the $^3S_1$-$^3D_1$ case still requires three counterterms
at subleading orders.

The counterargument is that, provided one regularizes first the EFT potential
and then regularizes and renormalizes the Schr\"odinger equation,
then for repulsive triplets the matrix elements of the primordial
counterterms will vanish as the cutoff is removed in the Schr\"odinger
equation (assuming the divergences in the potential are at worst
of the power-law type, which is the case).
The question is whether this observation about the primordial counterterms
is still correct when one simultaneously regularizes and renormalizes
the pion and nucleon loops.

Wilsonian renormalization as applied to nuclear EFT~\cite{Birse:1998dk,Birse:2005um,Valderrama:2014vra}
provides a different perspective into the problem of how to determine
the power counting.
There one imposes {\it renormalization group invariance} (RGI),
i.e. the condition that observables do not depend on the cutoff,
which for distorted wave perturbation theory at subleading
orders takes the form
\begin{eqnarray}
 \frac{d}{d R_c} \langle \Psi_{\rm LO} | V^{(\nu)} | \Psi_{\rm LO} \rangle = 0 \, ,
\end{eqnarray}
where $\Psi_{\rm LO}$ refers to the ${\rm LO}$ wave function.
If particularized to the contribution of the contact-range interactions, and
assuming a delta-shell regularization for simplicity, one arrives at
\begin{eqnarray}
  \frac{d}{d R_c} \left[ C(R_c) \frac{u_k^2(R_c)}{4 \pi R_c^2} \right] = 0 \, ,
\end{eqnarray}
which for the attractive and repulsive triplets might be approximated as
\begin{eqnarray}
  \frac{d}{d R_c} \left[\sin^2( 2 \sqrt{\frac{a_{3A}}{R_c}} + \phi)\,
    \frac{C_A(R_c)}{R_c^{1/2}} \right] &=& 0 \, , \\
  \frac{d}{d R_c} \left[e^{-4 \sqrt{\frac{a_{3R}}{R_c}}}\,
    \frac{C_R(R_c)}{R_c^{1/2}} \right] &=& 0 \, .
\end{eqnarray}
From the assumption that at $M R_c \sim 1$ the couplings are natural
($C \sim 1/M^2$), then evolving into $Q R_c \sim 1$:
\begin{eqnarray}
  C_A(R_c \sim \frac{1}{Q}) &\sim& \frac{1}{M^2}\,
  {\left( \frac{M}{Q} \right)}^{1/2} \, , \\
  C_R(R_c \sim \frac{1}{Q}) &\sim& \frac{1}{M^2}\,
  {\left( \frac{M}{Q} \right)}^{1/2}\,e^{-4 \sqrt{M\,a_{3R}}} \, ,
\end{eqnarray}
which implies that the contact-range couplings receive a power-law
enhancement of half an order over NDA.

Taking only into account the power-law enhancements, this means that
there are two counterterms at $\nu = 3/2$ (and three at $\nu = 7/2$),
while leading and subleading TPE enter at $\nu = 2$ and $3$.
Thus, from RGI the expectation is that these two counterterms will be able
to renormalize the scattering amplitudes of attractive and repulsive
uncoupled triplets up to $\nu < 7/2$, which is indeed the case~\cite{Valderrama:2009ei,Valderrama:2011mv,Long:2011qx,Long:2011xw}.

But in the repulsive case the couplings also display exponential
suppression at low energies.
Depending on how one deals with this exponential factor, one will reach
different conclusions about power counting:
\begin{itemize}
\item[(a)] If one concentrates on the power-law enhancements and suppressions
  but ignores the exponentials, the conclusion is that the $^3S_1$-$^3D_1$
  and $^3P_2$-$^3F_2$ channels require six counterterms each.
\item[(b)] On the contrary, if one considers that the exponential suppression
  completely supersedes the power-law enhancements, then one ends up with
  the two counterterms that have been shown to be enough to perturbatively
  renormalize the coupled triplets.
\end{itemize}
Indeed, the exponential suppression stemming from RGI provides an independent
explanation of why two counterterms can renormalize the coupled triplets:
Wilsonian renormalization does not make any assumption about the actual
separation of scales, and as $Q/M \to 0$ the couplings of repulsive
channels are arbitrarily suppressed and hence negligible
when compared with the couplings of attractive channels.
As a consequence calculations ignoring the counterterms in the repulsive
channels are expected to be cutoff independent.

Yet, there is a middle ground between the previous two interpretations of
the exponential demotion, which is that:
\begin{itemize}
\item[(c)] For a given value of the ratio $Q/M$, there is an equivalent
  power-law suppression
  \begin{eqnarray}
    e^{-4 \sqrt{M/Q}} \sim {\left( \frac{Q}{M} \right)}^{2 \nu_R} \, ,
  \end{eqnarray}
  with $\nu_R$ dependent on $Q/M$.
\end{itemize}
This is actually a compromise: it might not be an aesthetically
pleasing solution, but it is practical.
It has been tested in~\cite{Valderrama:2019yiv}, where I showed with a concrete
calculation that this type of demotion leads to a convergent EFT expansion
up to $\nu = 7$ (including higher order perturbation theory)
for the toy model presented in~\cite{Epelbaum:2018zli},
whose long-range part is singular and repulsive.
There the equivalent power-law suppression exponent
was taken to be $\nu_R = 3/4$, which leads to a demotion of
one order with respect to NDA for the contact-range couplings.
Though not explicitly discussed in~\cite{Valderrama:2019yiv},
not including this demotion worsens the convergence of
the EFT expansion.

Choice (a) and (b) actually appear as particular cases of equating the
exponential suppression to a specific power-law demotion.
If the separation of scales is poor ($Q/M \to 1$) the exponential factor
might be ignored ($\nu_R \to 0$), leading to no difference
in the power counting of the attractive and
repulsive channels.
On the contrary, if there is a perfect separation of scales ($Q/M \to 0$),
the equivalent power-law demotion of the repulsive channels
will be arbitrarily large ($\nu_R \to \infty$).
In between these two extremes there is the possibility of taking $\nu_R = 2$,
which would lead to three counterterms in the $^3S_1$-$^3D_1$ channel
(and one in the $^3P_1$ channel), making this choice equivalent to
the power counting originally advocated by Long and Yang~\cite{Long:2011xw}.

Even though the previous discussion has been limited to theoretical EFT
arguments, it is worth noticing that power counting can also be
potentially uncovered from the analysis of EFT truncation
errors, be it either within a Bayesian approach~\cite{Schindler:2008fh,Wesolowski:2015fqa,Furnstahl:2015rha,Melendez:2017phj,Wesolowski:2018lzj,Melendez:2019izc}, residual cutoff dependence~\cite{Griesshammer:2020fwr}, or by analyzing
the scaling properties of properly defined 
coefficients~\cite{Valderrama:2021bql}.
In this regard the eventual extension of Ref.~\cite{Thim:2023fnl} from
${\rm LO}$ to the different subleading power countings currently
available (with the specific intention of comparing which
is the best setup of contact-range couplings)
would be a particularly interesting way to determine
the power counting of coupled triplets.

To summarize, chiral leading and subleading TPE is perturbatively renormalizable
in the coupled triplets with two counterterms, a smaller number
than previously thought.
Whether power counting might require more counterterms -- either three as
in NDA or the calculations of Long and Yang~\cite{Long:2011xw},
or even six as in the renormalization group analysis
of Birse~\cite{Birse:2005um} --- is left here as an open question.
I have discussed a series of arguments and counterarguments, grounded
in theoretical EFT considerations, that tend to suggest that
a larger number of counterterms might be preferable.
Yet, it is worth noticing that besides the type of EFT ideas previously
discussed, there might exist more practical methods by which to elucidate
this question by analyzing the EFT truncation errors (though they lie
beyond the scope of the present manuscript).

\section*{Acknowledgments}

I would like to thank Bingwei Long and Chieh-Jen Yang for lively and
illuminating discussions regarding the results of this manuscript.
This work is partly supported by the National Natural Science Foundation
of China under Grant No. 12435007.


%

\end{document}